\documentstyle[prl,aps,preprint,tighten,floats,aps,amssymb]{revtex}
\input epsf.tex

\newcommand{\bmat}{\left(\begin{array}}
\newcommand{\emat}{\end{array}\right)}
\def\NPB#1#2#3{Nucl. Phys. B {\bf #1} (19#2) #3}
\def\PLB#1#2#3{Phys. Lett. B {\bf #1} (19#2) #3}

\def\PRD#1#2#3{Phys. Rev. D {\bf #1} (19#2) #3}

\def\yzero{\smash{\hbox{$y\kern-4pt\raise1pt\hbox{${}^\circ$}$}}}

\def\-{\hphantom{-}}
\def\ov{\overline}
\def\s2{\frac{1}{\sqrt2}}

\def\beq{\begin{equation}}
\def\eeq{\end{equation}}
\def\beqa{\begin{eqnarray}}
\def\eeqa{\end{eqnarray}}
\def\tr{{\rm tr \,}}
\def\Tr{{\rm Tr \,}}
\def\diag{{\rm diag \,}}

\def\IF{\relax{\rm I\kern-.18em F}}
\def\II{\relax{\rm I\kern-.18em I}}
\def\IP{\relax{\rm I\kern-.18em P}}

\def\Dsl{\,\raise.15ex\hbox{/}\mkern-13.5mu D} 

\def\IC{\bf C}
\def\IZ{\bf Z}
\def\IT{\bf T}
\def\z2z2{$\IC^3/(\IZ_2\times\IZ_2)$}

\def\id{{\bf 1}}

\newcommand{\drawsquare}[2]{\hbox{%
\rule{#2pt}{#1pt}\hskip-#2pt
\rule{#1pt}{#2pt}\hskip-#1pt
\rule[#1pt]{#1pt}{#2pt}}\rule[#1pt]{#2pt}{#2pt}\hskip-#2pt
\rule{#2pt}{#1pt}}

\newcommand{\fund}{\raisebox{-.5pt}{\drawsquare{6.5}{0.4}}}
\newcommand{\Ysymm}{\raisebox{-.5pt}{\drawsquare{6.5}{0.4}}\hskip-0.4pt%
        \raisebox{-.5pt}{\drawsquare{6.5}{0.4}}}
\newcommand{\Yasymm}{\raisebox{-3.5pt}{\drawsquare{6.5}{0.4}}\hskip-6.9pt%
        \raisebox{3pt}{\drawsquare{6.5}{0.4}}}
\newcommand{\antifund}{\overline{\fund}}

\def\s{\sigma}

\def\z{\zeta}

\def\bo{{\raise-.3ex\hbox{\large$\Box$}}}               
\def\face{{\raise.2ex\hbox{$\displaystyle \bigodot$}\mskip-2.2mu \llap {$\ddot
        \smile$}}}                                      

\def\leftrightarrowfill{$\mathsurround=0pt \mathord\leftarrow \mkern-6mu
        \cleaders\hbox{$\mkern-2mu \mathord- \mkern-2mu$}\hfill
        \mkern-6mu \mathord\rightarrow$}       
\def\dvec#1{\vbox{\ialign{##\crcr
        \leftrightarrowfill\crcr\noalign{\kern-1pt\nointerlineskip}
        $\hfil\displaystyle{#1}\hfil$\crcr}}}           

\def\beq{\begin{equation}}
\def\eeq{\end{equation}}

\def\beqx{\begin{displaymath}}
\def\eeqx{\end{displaymath}}

\def\beqa{\begin{eqnarray}}
\def\eeqa{\end{eqnarray}}

\begin{document}
\draft
\date{July 19, 2001}
\title{
\normalsize
\mbox{ }\hspace{\fill}
\begin{minipage}{10 cm}
UPR-943-T,
 CERN-TH/2001-182,
 CAMTP/01-8\\
{\tt hep-th/0107166}{\hfill}
\end{minipage}\\[5ex]
{\large\bf Chiral Four-Dimensional
N=1 Supersymmetric Type IIA Orientifolds
from  Intersecting D6-Branes
\\[1ex]}}

\author{ Mirjam Cveti\v c$^{1,2,3}$,  Gary Shiu$^{1,2}$ and Angel M. Uranga$^2$}
\address{$^1$Department of Physics and Astronomy, \\
University of Pennsylvania, Philadelphia PA 19104-6396, USA\\
 $^2$  Theory Division, CERN, CH-1211 Geneva 23, Switzerland\\
 $^3$Center for Applied Mathematics and Theoretical Physics,\\
 University of Maribor, SI-2000 Maribor, Slovenia}

\maketitle

\thispagestyle{empty}

\vspace*{-1cm}

\begin{abstract}

We construct $N=1$ supersymmetric four-dimensional orientifolds
of type IIA on $\IT^6/(\IZ_2\times \IZ_2)$ with D6-branes
intersecting at angles. The use of D6-branes not fully aligned
with the O6-planes in the model allows for a construction of many
supersymmetric models with chiral matter, including those with
the Standard Model and  grand unified gauge groups. We perform a
search for realistic gauge sectors, and construct the first
example of a supersymmetric type II orientifold with
$SU(3)_C\times SU(2)_L\times U(1)_Y$ gauge group and three
quark-lepton families. In addition to the
 supersymmetric Standard Model content, the model contains
right-handed neutrinos, a (chiral but anomaly-free) set of exotic
multiplets, and diverse vector-like multiplets. The general class
of these constructions  are related to familiar type II
orientifolds by small instanton transitions, which in some cases
change the number of generations, as discussed in specific
models. These constructions are supersymmetric only for special
choices of untwisted moduli. We briefly discuss the supersymmetry
breaking effects away from that point. The M-theory lift of this
general class of supersymmetric orientifold models should
correspond to purely geometrical backgrounds admitting a singular
$G_2$ holonomy metric and leading to four-dimensional M-theory
vacua with chiral fermions.
\end{abstract}
\newpage

\section{Introduction}

Four-dimensional $N=1$ supersymmetric Type II orientifolds
($\!$\cite{ABPSS,berkooz,N1orientifolds,zwart,ibanez,afiv,ShiuTye,wlm,CPW,kr,CUW}
and references therein) provide a class of consistent open-string
solutions which in turn could shed light on the physics of
strongly coupled heterotic string models. From a more
phenomenological viewpoint, they also provide a natural setup to
localized gauge interactions in lower dimensional sub-manifolds
of space-time, namely the D-branes, and lead to brane-world
models. Unfortunately, the constraints (arising from
Ramond-Ramond (RR) tadpole cancellation conditions) on
four-dimensional models are rather restrictive in the
supersymmetric case, and lead to relatively unrealistic gauge
sectors and matter contents in the simplest constructions.

Among the several discrete or continuous deformations of the simplest
models that have been explored, often mainly motivated by the search for
standard model like theories, we may recall

\begin{itemize}
\item
Blowing-up of orientifold singularities \cite{clew,pru}: The
resulting models are not described by a free world-sheet
conformal field theory (CFT), hence the space-time spectrum can
only be computed using field theory techniques. In compact
setups, the $\IZ_3$ orientifold of \cite{ABPSS} is the only
example in which this analysis has been performed \cite{clew} (it
is interesting to point out that results in D-brane models differ
in nature from those of perturbative heterotic orbifolds
\cite{C}).

\item
Locating the branes at different point in the internal
orbifold/orientifold space (see for instance \cite{wlm,cl,CUW}).
In a T-dual picture these possibilities correspond to the turning
of continuous or discrete Wilson lines. Explicit examples of
Wilson lines have been extensively considered, see for instance
\cite{afiv,wlm,CPW,lykken} for discrete Wilson lines, and
\cite{ibanez,afiv,cl} for continuous ones. For partially
successful attempts at supersymmetric model building using Wilson
lines see {\it e.g.},  \cite{CUW}.

\item
Introduction of discrete values for the
(Neveu-Schwarz-Neveu-Schwarz) NS-NS $B$ field
\cite{ShiuTye,sagnottietal}. The novel feature of this discrete
deformation is that the rank of the gauge group is reduced. This
can also be understood in the T-dual picture as tilting the tori,
thereby requiring fewer orientifold planes (see Appendix
\ref{tadpoles}) and hence fewer D-branes.

\item
Introduction of gauge fluxes in the D-brane world-volumes (see
\cite{bachas} for an early discussion). In the supersymmetric
context this has been explored in six-dimensional examples in
\cite{magnetised}. In a T-dual version, models with gauge fluxes
correspond to type II orientifolds with D-branes intersecting at
angles. Supersymmetric models in six and four dimensions
\cite{blumen6d,blumen4d,bonn1} have been constructed in the
situation with D6-branes parallel to the O6-planes in the model
\footnote{A non-compact version of these theories has been
studied in \cite{fhku}.}. Such constructions however lead to
non-chiral models. (In six dimensions, chiral supersymmetric
models with D-branes not parallel with the O-planes have been
obtained in \cite{bkl}, and implicitly in \cite{magnetised}).

\end{itemize}

Recently it has been realized that the RR tadpoles cancellation
conditions turn out to be much less constraining if one gives up
the requirement of supersymmetry. This observation in
\cite{ads,au} was exploited in the context of type IIB
orientifold model building in {\it e.g.},  \cite{aiq,aiqu,bailin}. It has
also allowed for a  construction of  type IIA orientifold and
orbifold models with D-branes at supersymmetry breaking angles and
realistic theories, with the Standard Model gauge group (or simple
extensions thereof) and three quark-lepton families
\cite{bkl,bgkl,afiru1,afiru2,bonn2,imr,bklo}
(for other proposals for realistic model-building using D-branes,
without a specific string construction or in a non-compact
set-up, see \cite{akt,aiqu,leigh}).

Despite this remarkable success, non-supersymmetric models have
more complicated dynamics than supersymmetric ones, hence are
less understood. In particular, tree level flat directions are
generically lifted by quantum corrections, leading to involved
stabilization problems. Also, the models contain uncancelled
NS-NS tadpoles which force to redefine the background geometry
\cite{nsns} \footnote{Some NS-NS tadpoles may be partially
avoided in models where the corresponding moduli are frozen to
discrete values, as in \cite{bklo}.}, as is obvious, for example,
from the existence of non-zero cosmological constant in the
models. For these and other reasons, it is more reassuring to
restrict to supersymmetric model building at the stringy level.
However, even for such models, the eventual supersymmetry
breaking required in any model attempting to describe realistic
low-energy physics, will lead to these or analogous omnipresent
issues, like the cosmological constant problem.

The purpose of this paper is explore the construction of four-dimensional
$N=1$ supersymmetric type IIA orientifolds with D6-branes intersecting at
angles, and leading to chiral gauge sectors. The simplest models satisfying
those requirements are orientifolds of toroidal type IIA orbifolds
$\IT^6/\IZ_N$ or $\IT^6/(\IZ_N\times \IZ_M)$, with D6-branes  not parallel
to the O6-planes. In this paper we focus on the $\IZ_2\times\IZ_2$
orbifold, for which the general pattern of the chiral spectrum is simple
enough. Extension to other orientifolds should be more involved, but
otherwise straightforward.

We succeeded in constructing the first $N=1$ supersymmetric model
with Standard Model gauge group and three quark-lepton families in
this setup \footnote{Notice the $N=1$ supersymmetric D3-brane
realistic model in Section 4.3 in \cite{aiqu} in a different
context.}. (The  letter version that summarizes the  key results
for this model appeared in \cite{csu}.)  Beyond the structure of
the minimal supersymmetric Standard Model (MSSM), the model
contains some additional gauge factors, right-handed neutrinos, a
chiral set of fields with exotic Standard Model gauge quantum
numbers, and diverse vector-like multiplets. Despite its lack of
fully realistic features, it provides the first construction of
phenomenologically appealing supersymmetric compactifications in
the setup of intersecting brane worlds. Moreover, a particularly
nice feature of such construction is that supersymmetry avoids
the hierarchy problem generically present in the (otherwise
realistic) models with D6-branes in
\cite{bgkl,afiru1,bkl,bonn2,imr}. Namely, our models will have a
relatively large string scale (close to the 4d Planck scale) and
not very large internal dimensions ($\ell\simeq ({\rm TeV})^{-1}$
at the largest).

We also discuss a number of interesting general results involved
in the construction. We describe the cancellation or previously
unnoticed mixed $U(1)$-gravitational anomalies, present in some
orientifold models (even without orbifold projection). They are
canceled by a Green-Schwarz mechanism mediated by untwisted RR
closed string fields, similar to that in \cite{afiru1} for mixed
non-Abelian anomalies.

The models under consideration, in a T-dual version, correspond
to chiral supersymmetric versions of the $\IZ_2\times \IZ_2$ type
IIB orientifold, with D9-branes with magnetic fluxes. We show
that such models are related to the familiar non-chiral
$\IZ_2\times \IZ_2$ orientifold in \cite{berkooz} by the T-dual
of a four-dimensional version of the small instanton transitions
\cite{small}. These transitions correspond, in the original
picture, to recombinations of D6-branes wrapped on intersecting
cycles. This provides a explicit picture of the transitions,
which allows to reproduce interesting phenomena. In particular,
we explicitly construct transitions where a toy model with
Standard Model-like gauge group changes the number of generations,
in a manner reminiscent of the chirality changing phase
transitions in \cite{ksafiuv,opp}.

Finally, the models upon consideration are supersymmetric only for
specific choices of the untwisted (complex structure) moduli.
Namely, the condition to preserve supersymmetry \cite{bdl} is
that the different D6-branes and O6-planes are related by
rotations in $SU(3)$. This implies certain constraints on the
angles among objects, in the three complex planes in $\IT^6$. For
fixed wrapped three-cycles, they imply constraints on the
untwisted complex structure moduli, so that supersymmetric
solutions exist generically only at isolated points in moduli
space. We briefly discuss the interesting question of the
dominant supersymmetry breaking effects upon small departures
from the supersymmetric points, and of supersymmetry restoration
by vacuum restabilization.

The paper is organized as follows. In Section \ref{rules} we
describe the construction of $\IZ_2\times \IZ_2$ orientifolds
with D6-branes at angles, and discuss their spectrum,
independently of supersymmetry. In Section \ref{cons} we
formulate the conditions to preserve 4d $N=1$ supersymmetry. We
present several explicit examples, including a four-family
Standard Model like theory, and a four-family $SU(5)$ grand
unified theory (GUT) model. The construction of three-family
models turns out to be very constrained, but we succeed in
building a model with SM gauge group (times additional factors)
and three quark-lepton families (plus additional exotic and
vector-like matter). The requirement of three families demands
introducing tilted complex structure in one two-torus, and
supersymmetry requires choosing specific ratios for the radii in
the remaining two-tori.

In Section \ref{small} we discuss how our chiral supersymmetric
models are related to the familiar non-chiral $\IZ_2\times \IZ_2$
orientifold \cite{berkooz}. In a T-dual version they are
connected through small instanton transitions. In the picture of
branes at angles the transition corresponds to a recombination of
3-cycles on which the D6-branes wrap. Such processes can mediate
phase transitions changing the number of chiral families, as we
illustrate in a toy construction with the Standard Model gauge
group. In Section \ref{ssb} we briefly discuss the supersymmetry
breaking effect in the open string sector when the ratios of
radii in the two-tori are chosen slightly away from the
supersymmetry-preserving values. These closed string moduli
couple as Fayet-Iliopoulos (FI) terms, hence they generate D-term
supersymmetry breaking terms whose  magnitude is related to a
deviation of the untwisted moduli away  from their  supersymmetric
values. Finally, in Section \ref{gtwo} we comment on more formal
applications of our constructions, which provide examples whose
lift to M-theory corresponds to compact 7-dimensional spaces
admitting $G_2$ holonomy metrics, and leading to chiral
four-dimensional gauge theories. Section \ref{conc} contains our
final comments.

\section{Model building rules for $\IZ_2\times \IZ_2$ orientifold with
branes at angles}

\label{rules}
In this Section we provide the rules to construct consistent orientifolds,
and to obtain the spectrum of massless states. We state these rules
independently of supersymmetry, so they are valid for non-supersymmetric
model building as well.

Our starting point is type IIA theory on $\IT^6/(\IZ_2\times \IZ_2)$,
with generators $\theta$, $\omega$ associated to the twists $v=(\frac
12,-\frac 12,0)$ and $w=(0,\frac 12,-\frac 12)$, hence acting as
\beqa
& \theta: & (z_1,z_2,z_3) \to (-z_1,-z_2,z_3) \nonumber \\
& \omega: & (z_1,z_2,z_3) \to (z_1,-z_2,-z_3)
\eeqa
where $z_i$ are complex coordinates in the $\IT^6$. For simplicity, we
consider the case of factorizable $\IT^6$.

We mod out this theory by the orientifold action $\Omega R$, where
$\Omega$ is world-sheet parity, and $R$ acts by
\beqa
R: (z_1,z_2,z_3) \to ({\ov z}_1,{\ov z}_2,{\ov z}_3)
\eeqa
The model contains four kinds of O6-planes, associated to the actions of
$\Omega R$, $\Omega R\theta$, $\Omega R \omega$, $\Omega R\theta\omega$,
as shown in Figure~\ref{orient}. We will not be specially interested in
the closed string sector, which can anyway be computed using standard
techniques. For the case of rectangular two-tori, it is as in
\cite{berkooz}, by T-duality.

In order to cancel the corresponding RR crosscap tadpoles, we introduce
D6-branes wrapped on three-cycles, which we consider factorized, namely
obtained as the product of one-cycles in each of the three two-tori.
Specifically, we consider $K$ stacks of $N_a$ D6-branes, $a=1,\ldots, K$,
wrapped on the $(n_a^i,m_a^i)$ cycle in the $i^{th}$ two-torus. We also
need to include the  images of these under the elements in the orientifold
group. Assuming for simplicity that our two-tori are rectangular
(extension to tilted two-tori is easy, and is discussed below), we include
$N_a$ D6-branes with wrapping numbers $(n_a^i,-m_a^i)$. For branes on top
of the O6-planes we also count branes and their images independently.

For future convenience, we define the homology class of the corresponding
3-cycles by
\beqa
[\Pi_a] = \prod_{i=1}^3 \,(n_a^i\, [a_i]\, +\, m_a^i\, [b_i])
\eeqa
and analogously for $[\Pi_{a'}]$. We also define the homology classes of
the cycles wrapped by the O6-planes, which for rectangular tori read
\beqa
[ \Pi_{\Omega R} ] = [ a_1 ] \times [ a_2 ] \times [ a_3 ] & \quad ; \quad
&
[ \Pi_{\Omega R \omega} ]= - [ a_1 ] \times [ b_2 ] \times [ b_3 ]
\nonumber
\eeqa
\beqa
[ \Pi_{\Omega R \theta\omega} ] = - [ b_1 ] \times [ a_2 ] \times [ b_3 ]
& \quad ; \quad &
[ \Pi_{\Omega R \theta} ] = - [ b_1 ] \times [ b_2 ] \times [a_3]
\label{defosix}
\eeqa
and we define $[\Pi_{O6}]=[\Pi_{\Omega R}] +[\Pi_{\Omega R \theta}]
+[\Pi_{\Omega R\omega}]+[\Pi_{\Omega R\theta\omega}]$.

Concerning the orbifold projections, let us focus on the case where the
branes pass through fixed points of the orbifold actions, hence $\theta$
and $\omega$ map each stack of branes to itself (and with the same 3-cycle
orientation). Extension to other cases is simple, and briefly mentioned in
Section IIIA. To specify the action of the different actions on the
Chan-Paton indices of the branes, for each stack of D6$_a$-branes, and
their $\Omega R$ images, denoted D6$_{a'}$-branes, we introduce the
Chan-Paton actions
\beqa
\gamma_{\theta,a} & = & \diag(i \id_{N_a/2},-i \id_{N_a/2}\, ;
-i \id_{N_a/2},i \id_{N_a/2}) \nonumber \\
\nonumber \\
\gamma_{\omega,a} & = & \diag \left[
\pmatrix{0 & \id_{N_a/2} \cr -\id_{N_a/2} & 0 } \; ; \;
\pmatrix{0 & \id_{N_a/2} \cr -\id_{N_a/2} & 0 } \right] \nonumber \\
\nonumber \\
\gamma_{\Omega R,a} & = &
\pmatrix{ & & \id_{N_a/2} & 0 \cr
& & 0 & \id_{N_a/2} \cr
\id_{N_a/2} & 0 & & \cr
0 & \id_{N_a/2} & & \cr }
\eeqa
The actions for the orbifold group form a projective representation, which
corresponds to the choice of closed string sector usually known as
without discrete torsion \footnote{Our D6-branes wrap special lagrangian
3-cycles (A-branes), hence they carry projective representations in the model
without discrete torsion \cite{gaberdiel}. In a T-dual (mirror) version,
the system is mapped to a set of D9-branes with holomorphic bundles
(B-branes) in a model with discrete torsion, which again carry projective
representations \cite{douglas} (for orientifolds with discrete torsion,
see \cite{krdiscrete}).}.

The models are constrained by RR tadpole cancellation conditions.
Orientifolds by $\Omega R$ action do not generate twisted crosscaps
\cite{blumen6d,blumen4d,bonn1}, hence the twisted disk tadpoles should
vanish. The simplest way to accomplish this is to choose traceless
Chan-Paton matrices, as done above\footnote{As pointed out in \cite{bkl},
and implicitly used in \cite{magnetised}, it is possible to achieve this
for non-traceless choices if for any point fixed under the orientifold and
some orbifold action the total Chan-Paton trace for the different branes
passing through it cancels. We will not consider this case in the present
analysis.}. Cancellation of untwisted RR tadpoles simply requires the
cancellation of D6-brane and O6-plane charge, namely
\beqa
\sum_{a} N_a\, [\Pi_a] + \sum_a N_a\, [\Pi_{a'}] + (-4)\times 8\,
[\Pi_{O6}] = 0
\label{homosum}
\eeqa
It is further discussed in Appendix A, and leads to the constraints
(\ref{tadpole1}).

\medskip

Such choices define a consistent model, for which the resulting spectrum
is discussed in the following. The results for branes at generic angles are
shown in Table~\ref{matter}.

Let us consider the $aa$ sector (strings stretched within a single stack
of D6$_a$-branes) which is invariant under $\theta$, $\omega$, and which
is exchanged with $a'a'$ by the action of $\Omega R$. For the gauge group,
the $\theta$ projection breaks $U(N_a)$ to $U(N_a/2)\times U(N_a/2)$, and
the further $\omega$ projection identifies both factors, leaving $U(N_a/2)$.
Concerning the matter multiplets, we obtain three adjoint $N=1$ chiral
multiplets. This sector is however not $N=4$ since the superpotential for
the adjoints $\Phi_1$, $\Phi_2$ and $\Phi_3$ reads
\beqa
W=\Tr (\Phi_1 \Phi_2 \Phi_3 + \Phi_1\Phi_3\Phi_2)
\eeqa
instead of the $N=4$ commutator structure. This agrees with the result in
\cite{douglas} in the T-dual (mirror) picture. For branes parallel to some
O6-plane the projections are identical to \cite{berkooz}, leading to a
$USp(N_a)$ group with three $N=1$ chiral multiplets in the
two-index antisymmetric representation (in our `antisymmetric' of
symplectic factors we also include the singlet).

The $ab+ba$ sector, strings stretched between D6$_a$- and D6$_b$-branes,
is invariant as a whole under the orbifold projections, and is mapped to
the $b'a'+a'b'$ sector by $\Omega R$. The matter content before any
projection would be given by $I_{ab}$ chiral
fermions in the bifundamental $(N_a,{\ov N}_b)$, where
\beqa
I_{ab}=[\Pi_a]\cdot [\Pi_b]=\prod_{i=1}^3 (n_a^i m_b^i - m_a^i n_b^i)
\eeqa
is the intersection number of the wrapped cycles, and the sign of $I_{ab}$
denotes the chirality of the corresponding fermion (our default convention
is that negative intersection numbers correspond to left-handed
fermions). For supersymmetric intersections, additional massless scalars
complete the corresponding supermultiplet. For non-supersymmetric
intersections, the masses for light scalars are as in \cite{afiru1}.

In principle, in performing the orbifold quotient one needs to take into
account the orbifold action on the intersection point. The final result
however turns out to be rather insensitive to this subtlety, as opposed
to the six-dimensional case \cite{bkl}. For an intersection point fixed
under $\theta$ and $\omega$, the orbifold projections reduce the matter
content to a bifundamental $(\fund_a,\antifund_b)$ of $U(N_a/2)\times
U(N_b/2)$. For intersection points exchanged by the orbifold actions, for
instance two points fixed under $\omega$ but exchanged under $\theta$, we
should consider just one point (the other being merely its $\theta$-image)
and not impose the $\theta$ projection. The resulting fields are {\em two}
bifundamentals. Due to this compensation, the total number of fields in
the $ab$ sector is simply $I_{ab}$ chiral fermions in the
$(\fund_a,\antifund_b)$ of $U(N_a/2)\times U(N_b/2)$ (plus scalars, which
fill out supermultiplets in the supersymmetric case).

A similar effect takes place in $ab'+b'a$ sector, for $a\neq b$, where
the
final matter content is given by $I_{ab'}$ chiral fermions in the
bifundamental $(\fund_a,\fund_b)$.

Finally, let us consider the $aa'+a'a$ sector. In this case, the orbifold
action on the intersection point turns out to be crucial. At an intersection
point with angle $\theta_i$ in the $i^{th}$ two-torus states are labeled,
in the bosonized formulation (see \cite{afiru1}), by a vector
$(r_1+\theta_1,r_2+\theta_2,r_3+\theta_3,r_4)$, where $r_i=\IZ,\IZ+\frac
12$ in the NS, R sectors, respectively,
and $\sum_i r_i={\rm odd}$ due to the GSO
projection, and $r_4=-1/2, +1/2$ corresponds to respective
left-handed and right-handed
fermions (in our default convention). For an intersection point invariant
under $\theta$ and $\omega$, the eigenvalues of such state under $\theta$,
$\omega$ and $\Omega R$ are $\exp(2\pi i\, r\cdot v)$, $\exp(2\pi i\,
r\cdot w)$, and $-1$, where recall $v=(1,-1,0,0)/2$, $w=(0,1,-1,0)/2$ are
the twist vectors. The projections on the Chan-Paton factors are
\beqa
\lambda & = & e^{2\pi i\, rv} \gamma_{\theta,6} \lambda
\gamma_{\theta,6'}^{-1} \nonumber \\
\lambda & = & e^{2\pi i\, rw} \gamma_{\theta,6} \lambda
\gamma_{\theta,6'}^{-1} \nonumber \\
\lambda & = & - \gamma_{\Omega R} \lambda^T \gamma_{\Omega R}^{-1}
\eeqa
Before the orientifold projection, one gets a chiral fermion in the
bifundamental $(N_a/2,{\ov {N_a/2}}\,')$, regardless of the $\theta$,
$\omega$ eigenvalues of the state. The orientifold projection, however,
distinguishes the different cases and leads to a two-index antisymmetric
representation of $U(N_a/2)$, except for states with $\theta$ and $\omega$
eigenvalue $+1$, where it yields a two-index symmetric representation.

Now consider points not fixed under some orbifold element, say two
points fixed under $\omega$, and exchanged by $\theta$. Then one simply
keeps one point, and does not impose the $\omega$ projection.
Equivalently, one considers all possible eigenvalues for $\omega$, and
applies the above rule to read off whether the symmetric or the
antisymmetric representation survives.

It is possible to give a closed formula for the precise chiral
matter content in this sector, which basically follows from
cancellation of non-Abelian anomalies. The final result for the
net number of symmetric and antisymmetric representations in the
$aa'$ sector is \beqa n_{\Ysymm} & = & -\frac 12 (I_{aa'} -
\frac{4}{2^{k}} I_{a,O6}) \nonumber
\\
n_{\Yasymm} & = & -\frac 12 (I_{aa'} + \frac{4}{2^{k}} I_{a,O6})
\label{aaprime}
\eeqa
where $k$ is the number of tilted tori and $I_{aa'}=[\Pi_a]\cdot
[\Pi_a']$, $I_{a,O6}=[\Pi_a]\cdot[\Pi_{O6}]$.
Notice that the definitions (\ref{defosix}) should be modified in the
obvious way for tilted tori. The result (\ref{aaprime}) is then general
\footnote{A sketch of the anomaly argument is as follows. Tadpole
cancellation (\ref{homosum}) implies that $\sum_b N_b I_{ab} + \sum_{b}
N_b I_{ab'} - \frac {32}{2^k} I_{a,O6} =0$. The first two members, for
$b\neq a$ give (minus two times) the $SU(N_a/2)$ anomaly due to $ab$ and
$ab'$ fields. The rest, which equals $N_a I_{aa'} - \frac {32}{2^k}
I_{a,O6}$ must correspond to (minus two times) the anomaly in the
$aa'$ sector, which is given by $(-2)\times (n_{\Ysymm} (N_a/2+4) +
n_{\Yasymm} (N_a/2-4))$. Equating both for arbitrary $N_a$ yields
(\ref{aaprime}).}.
Clearly, getting the non-chiral piece requires the full computation of the
spectrum.

\begin{table}[htb] \footnotesize
\renewcommand{\arraystretch}{1.25}
\begin{center}
\begin{tabular}{|c|r|}
\hspace{2cm} {\bf Sector} \hspace{2cm} &
{\bf Representation} \hspace{4cm} \\
\hline\hline
$aa$   & $U(N_a/2)$ vector multiplet \hspace{3.3cm} \\
       & 3 Adj. chiral multiplets  \hspace{3.3cm} \\
\hline\hline
$ab+ba$   & $I_{ab}$ $(\fund_a,\antifund_b)$ fermions  \hspace{3.3cm} \\
\hline\hline
$ab'+b'a$ & $I_{ab'}$ $(\fund_a,\fund_b)$ fermions \hspace{3.3cm}  \\
\hline\hline
$aa'+a'a$ & $-\frac 12 (I_{aa'} - \frac{4}{2^{k}} I_{a,O6}) \;\;
\Ysymm\;\;$ fermions  \hspace{3.3cm} \\
          & $-\frac 12 (I_{aa'} + \frac{4}{2^{k}} I_{a,O6}) \;\;
\Yasymm\;\;$ fermions \hspace{3.3cm}
\end{tabular}
\end{center}
\caption{\small General spectrum on D6-branes at generic angles
(namely, not parallel to any O6-plane in all three tori). The spectrum is
valid for tilted tori. The models may contain additional non-chiral pieces
in the $aa'$ sector and in $ab$, $ab'$ sectors with zero intersection, if
the relevant branes overlap. In supersymmetric situations, scalars combine
with the fermions given above to form chiral supermultiplets.
\label{matter} }
\end{table}

\section{Construction of supersymmetric models}
\label{cons}

In this Section we turn to the construction of $N=1$
supersymmetric models. The condition that the system of branes
preserve the $N=1$ supersymmetry unbroken in the closed string
sector amounts, following \cite{bdl}, to requiring that each
stack of D6-branes is related to the O6-planes by a rotation in
$SU(3)$. More specifically, denoting by $\theta_i$ the angles the
D6-brane forms with the horizontal direction in the $i^{th}$
two-torus, supersymmetry preserving configurations must satisfy
\beqa \theta_1\, +\, \theta_2\, +\, \theta_3\, =\, 0 \eeqa For
fixed wrapping numbers $(n^i,m^i)$, the condition translates into
a constraint on the ratio of the two radii on each torus. For
rectangular tori, denoting $\chi_i=(R_2/R_1)_i$, with $R_2, R_1$
the vertical resp. horizontal directions, the constraint is
\footnote{Notice that an overall rescaling of any two-torus
leaves the conditions unaffected. This agrees with the expected
property that supersymmetric features of A-branes are independent
of K\"ahler moduli, and depend on complex structure moduli alone.}
\beqa \arctan (\chi_1 \frac{m_1}{n_1})\, +\, \arctan (\chi_2
\frac{m_2}{n_2})\, +\, \arctan (\chi_3 \frac{m_3}{n_3})\, =\, 0
\eeqa The modification for tilted tori is straightforward and
will be discussed later.

It is possible to find sets of D6-branes solving the tadpole conditions
and preserving supersymmetry for some choice of $\chi_i$. A simple example
with non-trivial angles in all three tori is provided in
Section~\ref{gut}. Clearly, as the number of branes at angles increases,
the constraints to preserve supersymmetry get more involved and there may
not exist solutions to the $\chi_i$ for a given set of RR tadpole-free
wrapping numbers.

In order to simplify the supersymmetry conditions, and our search
for realistic models, we will consider a particular Ansatz for the
kind of D6-branes in our configuration. We will consider that the
D6-branes, not parallel to any O6-plane, will have angles (with
respect to the O6-plane)  of the form $(\theta_1,\theta_2,0)$,
$(\theta_1,0,\theta_3)$ or $(0,\theta_2,\theta_3)$. Such an angle
structure makes the supersymmetry conditions relatively simple.
An additional advantage is that having the branes parallel to
some O6-plane helps in avoiding chiral matter in $aa'$ sectors
(although not completely in all cases). It is interesting to
observe that, despite the fact that our brane system is the
composition of three different ``six-dimensional'' configurations,
the Ansatz is rich enough to allow for a construction of chiral
models, and moreover yields realistic structures, as we show in
the remainder of this Section. Further exploration of more general
models is left for future research. In the following, we turn to
the construction of several explicit,  potentially
phenomenologically viable models.

\subsection{Four-family model}

The general structure of the chiral spectrum in the $\IZ_2\times
\IZ_2$ models is relatively similar to that of the
``un-orbifolded'' models. A simple consequence of this observation
for the construction of standard model like theories is that,
following the argument in \cite{bgkl}, the net number of
left-handed quarks is even if all tori are rectangular
\footnote{This assumes that $U(3)$ and $U(2)$ arise from generic
D6-branes. Another possibility would be that avoided the $SU(2)$
factor is obtained as $USp(2)$ from D6-branes on top of
O6-planes.}.

\begin{table}[htb] \footnotesize
\renewcommand{\arraystretch}{1.25}
\begin{center}
\begin{tabular}{|c||c|l|}
Type & $N_a$ & $(n_a^1,m_a^1) \times (n_a^2,m_a^2) \times (n_a^3,m_a^3)$ \\
\hline
$A_1$ & 6+2 & $(1,1)\times(1,-2)\times (1,0)$ \\
$A_2$ & 2 & $(1,0) \times(1,0) \times (1,0)$ \\
\hline
$B_1$ & 4 & $(1,0) \times (1,2) \times (1,-1)$ \\
$B_2$ & 8 & $(1,0) \times (0,1) \times (0,-1)$ \\
\hline
$C_1$ & 2 & $(1,2) \times (1,0) \times (1,-2)$ \\
$C_2$ & 8 & $(0,1) \times (1,0) \times (0,-1)$ \\
\end{tabular}
\end{center}
\caption{\small D6-brane configuration for the four-family model}
\label{fourgen}
\end{table}

\begin{table} 
\footnotesize
\renewcommand{\arraystretch}{1.25}
\begin{center}
\begin{tabular}{|c||c||c|c|c|c||c|c|}
Sector & $U(3) \times U(2) \times USp(2) \times USp(8)\times USp(8)$ &
$Q_3$ & $Q_1$ & $Q_2$ & $Q_1'$ & $Q_Y$ & Field \\
\hline
$A_1 B_1$ & $4\times (3,{\ov 2},1,1,1)$ &
$1$ & $0$ & $-1$ & $0$ & $-\frac 16$ & $Q_L$ \\
          & $4\times (1,{\ov 2},1,1,1)$ &
$0$ & $1$ & $-1$ & $0$ & $-\frac 12$ & $L$ \\
$A_1 B_2$ & $(3,1,1,8,1)$ &
$1$ & $0$ & $0$ & $0$ & $-\frac 13,\frac 23$ & ${\ov U}, {\ov D}$ \\
          & $(1,1,1,8,1)$ &
$0$ & $1$ & $0$ & $0$ & $0,-1$ & ${\ov \nu}, {\ov E}$ \\
$A_1 C_1$ & $4\times (\overline{3},1,1,1,1)$ &
$-1$ & $0$ & $0$ & $1$ & $-\frac 16$ & \\
          & $4\times (1,1,1,1,1)$ &
$0$ & $-1$ & $0$ & $1$ & $\frac 12$ & \\
$A_1 C_2$ & $2 \times (\ov{3},1,1,1,8)$ &
$-1$ & $0$ & $0$ & $0$ & $\frac 13,-\frac 23$ & $U, D$ \\
          & $2 \times (1,1,1,1,8)$ &
$0$ & $-1$ & $0$ & $0$ & $0,1$ & $\nu, E$ \\
$B_1 C_1$ & $4\times (1,2,1,1,1)$ &
$0$ & $0$ & $1$ & $-1$ & $0$ &  \\
$B_1 C_2$ & $2 \times (1,2,1,1,8)$ &
$0$ & $0$ & $1$ & $0$ & $\pm \frac 12$ & $H_U, H_D$  \\
$B_2 C_1$ & $2 \times (1,1,1,8,1)$ &
$0$ & $0$ & $0$ & $1$ & $\pm \frac 12$ & $H_U, H_D$  \\
\hline
$A_1 C_1^{\prime}$ & $12 \times (\ov{3},1,1,1,1)$ &
$-1$ & $0$ & $0$ & $-1$ & $-\frac 16$ &  \\
                   & $12 \times (1,1,1,1,1)$ &
$0$ & $-1$ & $0$ & $-1$ & $\frac 12$ &  \\
$B_1 C_1^{\prime}$ & $12 \times (1,2,1,1,1)$ &
$0$ & $0$ & $1$ & $1$ & $0$ &  \\
\hline
$A_1 A_1^{\prime}$ & $2\times (3,1,1,1,1)$ &
$-2$ & $0$ & $0$ & $0$ & $-\frac 13$ &  \\
                   & $2\times (6,1,1,1,1)$ &
$2$ & $0$ & $0$ & $0$ & $\frac 13$ &  \\
                   & $2\times (1,1,1,1,1)$ &
$0$ & $2$ & $0$ & $0$ & $1$ &  \\

$B_1 B_1^{\prime}$ & $2 \times (1,1,1,1,1)$ &
$0$ & $0$ & $2$ & $0$ & $0$ &  \\
                   & $2 \times (1,\ov{3},1,1,1)$ &
$0$ & $0$ & $-2$ & $0$ & $0$ &  \\
\end{tabular}
\end{center}
\caption{\small Chiral open string spectrum for the four-family model. For
convenience we have changed our default convention, so that positive
intersections give left handed fermions.
The $U(1)$ are the overall $U(1)$ motion of the $A_1$ branes (split into
groups of $6$ and $2$), $B_1$ and $C_1$ branes respectively. For clarity,
we have not listed the $aa$ sectors. We distinguish the $2$ and the ${\ov
2}$ of $U(2)$ in order to keep track of their $U(1)$ charges.
\label{spectrum4}}
\end{table}

In this Section, we consider the case of rectangular tori, as
an illustration and warm-up, and present a model with the Standard Model
like group and four generations. The model we present is by no means
unique; one can construct rather straightforwardly other four-family
models along the lines we discussed here.
Let us consider branes wrapping
around the 3-cycles shown in Table~\ref{fourgen}. The model
preserves supersymmetry if $\chi_1=2\chi_2=\chi_3$.

The $8$ D6-branes labeled $A_1$ are split into two parallel but not
overlapping stacks of $6$ and $2$ branes, giving rise to $U(3) \times
U(1)$ gauge group. This can be achieved by placing them at different
positions in one of the two-tori.

Note that one of these two $U(1)$'s, namely $Q_3-3 Q_1$, is
actually a generator within the $SU(4)$ part of the $U(4)$ when
the $6+2$ branes coincide. This ensures that this $U(1)$ is
non-anomalous (see \cite{afiru1,imr} for $U(1)$ anomalies in this
context), and moreover do not have the linear $B\wedge F$
couplings which make some of the non-anomalous $U(1)$ massive (as
pointed out in \cite{imr}). We will make use of this fact in
seeking for the hyper-charge in the model.

The chiral open string spectrum is tabulated in Table \ref{spectrum4}.
For clarity, we have displayed the spectrum when the D6-branes
labeled $B_1$, $B_2$, $C_2$ are on top of some O6-planes.
Therefore, the corresponding gauge groups are $USp(4)$, $USp(8)$ and
$USp(8)$ respectively. In a more generic situation, we can move these
$D6$-branes (and their images under $\theta$, $\omega$ and $\Omega R$)
away from the O6-planes, in a way consistent with the orbifold and
orientifold projections. The basic unit that can be moved away from the
O6-planes is $4$ D6-branes and their $4$ orientifold images. In the
effective field theory, the generic model corresponds to Higgsing of the
$USp (4n)$ group to $U(1)^n$, and decomposing the fundamental of $USp(4n)$
as
\begin{equation}
{\bf 4n} =
2 \times (\pm 1,0,\dots,0) + 2 \times (0,\pm1,\dots,0)
+ \dots 2 \times (0,\dots,0,\pm 1)
\end{equation}
We will actually be interested in the generic model when these
$D6$-branes are away from the O6-planes, in order to generate
additional $U(1)$'s to enter the hyper-charge generator. The gauge
group is broken to $SU(3) \times SU(2)$ plus a number of $U(1)$
factors. Let $Q_4$, $Q_{8B}$, $Q_{8B}'$, $Q_{8C}$ and $Q_{8C}'$
be the $U(1)$ generators arising from moving the $B_1$, $B_2$ and
$C_2$ branes away from the O6-planes respectively. It is easy to
check that these $U(1)$'s are automatically non-anomalous and
massless, consistent with the fact that they arise from some
non-Abelian gauge groups. Hence, we can form the non-anomalous and
massless linear combination
\begin{equation}
Q_Y = \frac{1}{6} Q_3 - \frac{1}{2} Q_1
+ \frac{1}{2} \left( Q_{8B} + Q_{8B}' \right)
+ \frac{1}{2} \left( Q_{8C} + Q_{8C}' \right)
\label{hypy}
\end{equation}
The remaining $U(1)$'s (except those arising from non-Abelian
generators) have $B \wedge F$ couplings and become massive. The spectrum
for this generic model is easily obtained from that in the Table
by splitting the fundamental representations of the symplectic
factors as explained above.

The charge (\ref{hypy}) gives a good candidate for hyper-charge
in this model, as can be seen from the charge assignments for
different fields, shown in the Table. In particular, the
left-handed quarks and leptons come from $A_1B_1$. Four net
families of up and down anti-quarks and right-handed leptons and
neutrinos, along with some vector-like pairs, arise from the $A_1
C_2$ and $A_1 B_2$ sectors. There are many candidates for the
Higgs fields in the  supersymmetric Standard Model, from the $B_1
C_2$ and $B_2 C_1$ sectors. In addition, there are exotic chiral
(but anomaly-free) sets of fields, plus vector-like multiplets
under the Standard Model.

In the Table above we have not computed the non-chiral pieces of
the $ab$, $ab'$ and $aa'$ spectrum, because we can get rid of
them by a simple mechanism. Non-chiral matter arises whenever
two-branes are parallel in some complex plane. Locating them at
different positions in that plane makes such a non-chiral matter
massive. This can be done consistently with the orbifold
projection, as follows. Consider a stack of $2N$ branes mapped to
itself under $\theta$ and $\omega$. Moving them off in {\it e.g.},
the first plane amounts to splitting them in two stacks of $N$
branes, fixed under $\omega$ and exchanged by $\theta$. The choice
$\gamma_{\omega}=i\id_N$ for one stack and
$\gamma_{\omega}=-i\id_N$ for the other is consistent, and leads
to  exactly the same chiral spectrum as before the motion. This
is not surprising since the deformation is continuous and cannot
modify the chiral structure. From the effective field theory
viewpoint, this amounts to turning on a nonzero vacuum
expectation value (vev) for the singlet part in the $U(N)$ adjoint
multiplets in the $aa$ sector; the non-chiral multiplets get
massive due to non-trivial superpotentials, which are easy to
obtain although we do not discuss them explicitly.

In the construction above, and the subsequent ones we will assume we are
performing this operation to avoid phenomenologically undesirable
non-chiral matter.

\subsection{Three-family model}

Following \cite{bkl}, the even generation number problem can be solved by
considering some tori to be tilted. As discussed in \cite{ab} the tilting
parameter is discrete by the orientifold symmetry, and it can take only
one non-trivial value. This mildly modifies the closed string sector, and
in particular implies the existence of fewer O6-planes in the model.
Concerning the D6-brane sector, a 1-cycle $(n_a^i,m_a^i)$ along a tilted
torus is mapped to $(n^i,-m^i-n^i)$. It is convenient to define
${\widetilde m}^i=m^i-\frac 12 n^i$, and label the cycles as
$(n^i,{\widetilde m}^i)$. The tadpole cancellation conditions are computed
in Appendix A for the case of tilting just the third two-torus, and lead
to (\ref{tadpole2}), which are naturally expressed in terms of the
redefined labels. It is also easy to check that intersection numbers are
\beqa
I_{ab} =
(n_a^1 m_b^1 - n_b^1 m_a^1) \times
(n_a^2 m_b^2 - n_b^2 m_a^2) \times
(n_a^3 {\widetilde m}_b^3 - n_b^3 {\widetilde m}_a^3)
\eeqa
Due to the smaller number of O6-planes in tilted configurations,
RR tadpole conditions are very constraining for more than one
tilted torus. Centering on tilting just the third torus, the
search for theories with $U(3)$ and $U(2)$ gauge factors carried
by branes at angles and three left-handed quarks, turns out to be
very constraining, at least within our Ansatz. We have found
essentially a unique solution, with D6-brane configuration with
wrapping numbers $(n_a^i,\widetilde{m}_a^i)$ given in Table
\ref{cycles3family}. The configuration is supersymmetric for
$\chi_1:\chi_2:\chi_3=1:3:2$.

\begin{table} 
[htb] \footnotesize
\renewcommand{\arraystretch}{1.25}
\begin{center}
\begin{tabular}{|c||c|l|}
Type & $N_a$ & $(n_a^1,m_a^1) \times
(n_a^2,m_a^2) \times (n_a^3,\widetilde{m}_a^3)$ \\
\hline
$A_1$ & 8 & $(0,1)\times(0,-1)\times (2,{\widetilde 0})$ \\
$A_2$ & 2 & $(1,0) \times(1,0) \times (2,{\widetilde 0})$ \\
\hline
$B_1$ & 4 & $(1,0) \times (1,-1) \times (1,{\widetilde {3/2}})$ \\
$B_2$ & 2 & $(1,0) \times (0,1) \times (0,{\widetilde {-1}})$ \\
\hline
$C_1$ & 6+2 & $(1,-1) \times (1,0) \times (1,{\widetilde{1/2}})$ \\
$C_2$ & 4 & $(0,1) \times (1,0) \times (0,{\widetilde{-1}})$ \\
\end{tabular}
\end{center}
\caption{\small D6-brane configuration for the three-family model.}
\label{cycles3family}
\end{table}


\begin{table} \footnotesize
\renewcommand{\arraystretch}{1.25}
\begin{center}
\begin{tabular}{|c||c||c|c|c|c|c||c|c|c|}
Sector & $U(3)\times U(2)\times USp(2)\times USp(2)\times USp(4)$ &
$Q_3$ & $Q_1$ & $Q_2$ & $Q_8$ & $Q_8'$ & $Q_Y$ & $Q_8-Q_8'$ & Field
\\
\hline
$A_1 B_1$ & $3 \times 2\times (1,{\ov 2},1,1,1)$ &
0 & 0 & $-1$ & $\pm 1$ & 0 & $\pm \frac 12$ & $\pm 1$ & $H_U$, $H_D$\\
          & $3\times 2\times (1,{\ov 2},1,1,1)$ &
0 & 0 & $-1$ & 0 & $\pm 1$ & $\pm \frac 12$ & $\mp 1$ & $H_U$, $H_D$\\
$A_1 C_1$ & $2 \times (\ov{3},1,1,1,1)$ &
$-1$ & 0 & 0 & $\pm 1$ & 0 & $\frac 13, -\frac 23$ & $1,-1$ & $U$, $D$\\
          & $2 \times (\ov{3},1,1,1,1)$ &
$-1$ & 0 & 0 & 0 & $\pm 1$ & $\frac 13, -\frac 23$ & $-1,1$ & $U$, $D$\\
          & $2 \times (1,1,1,1,1)$ &
0 & $-1$ & 0 & $\pm 1$ & 0 & $1,0$ & $1,-1$ & $E$, $\nu_R$\\
          & $2 \times (1,1,1,1,1)$ &
0 & $-1$ & 0 & 0 & $\pm 1$ & $1,0$ & $-1,1$ & $E$, $\nu_R$\\
$B_1 C_1$ & $(3,{\ov 2},1,1,1)$ &
1 & 0 & $-1$ & 0 & 0 & $\frac 16$ & 0 & $Q_L$\\
             & $(1,{\ov 2},1,1,1)$ &
0 & 1 & $-1$ & 0 & 0 & $-\frac 12$ & 0 & $L$\\
$B_1 C_2$ & $(1,2,1,1,4)$ &
0 & 0 & $1$ & 0 & 0 & 0 & 0 & \\
$B_2 C_1$ & $(3,1,2,1,1)$ &
1 & 0 & 0 & 0 & 0 & $\frac 16$ & 0 & \\
          & $(1,1,2,1,1)$ &
0 & 1 & 0 & 0 & 0 & $-\frac 12$ & 0 & \\
$B_1 C_1^{\prime}$ & $2\times (3,2,1,1,1)$ &
1 & 0 & 1 & 0 & 0 & $\frac 16$ & 0 & $Q_L$ \\
                   & $2\times (1,2,1,1,1)$ &
0 & 1 & 1 & 0 & 0 & $-\frac 12$ & 0 & $L$ \\
\hline
$B_1 B_1^{\prime}$ & $2\times (1,1,1,1,1)$ &
0 & 0 & $-2$ & 0 & 0 & 0 & 0 &  \\
                   & $2\times (1,3,1,1,1)$ &
0 & 0 & $2$ & 0 & 0 & 0 & 0 &
%
\end{tabular}
\end{center}
\caption{\small Chiral Spectrum of the open string sector in the
three-family model. Notice that we have not included the $aa$ sector, 
even though it is generically present in the model. As explained in
the text, the non-chiral pieces in the $ab$, $ab'$ and $aa'$ sectors are
generically not present.
\label{spectrum3}}
\end{table}

The $8$ D6-branes labeled $C_1$ are spit in two parallel but not
overlapping stacks of $6$ and $2$ branes, hence lead to a gauge group
$U(3)\times U(1)$. For simplicity we may choose them to pass through
different $\IZ_2$ fixed points in some two-torus (alternatively, we may
locate them at generic positions in one two-torus, as described in the
previous Section).

It is interesting to observe that one of these two $U(1)$'s is
actually a generator within the $SU$ part of the $U(4)$
gauge group that would arise for coincident branes.
This feature ensures that this $U(1)$ is automatically
non-anomalous, and moreover does not have $B\wedge F$ couplings.
This $U(1)$ will turn out to be crucial in the appearance of
hyper-charge in this model.

For convenience it is also useful to consider the $8$ D6-branes
labeled $A_1$ to be away from the O6-planes in all three complex
planes. One is left with two D6-branes that can move
independently (hence give rise to a group $U(1)^2$), plus their
$\theta$, $\omega$ and $\Omega R$ images. The spectrum is
computed applying standard rules, even though they differ
slightly from our explicit rules above. In the effective theory,
the generic model corresponds to a ``Higgsing'' of $USp(8)$ down
to $U(1)^2$. A possibility to get the chiral spectrum below, with
a minimum amount of effort is to merely decompose the chiral
spectrum for D6-branes on top of the O6-plane, with respect to
the surviving group $U(1)^2$.

The open string spectrum is tabulated in Table \ref{spectrum3}.
The generators $Q_3$, $Q_1$ and $Q_2$ refer to the $U(1)$ factor
within the corresponding $U(n)$, while $Q_8$, $Q_8'$ are the
$U(1)$'s arising from the $USp(8)$. The next two columns provide
the charges under the only non-anomalous $U(1)$'s, namely
$Q_8-Q_8'$ and \beqa Q_Y & = & \frac 16 Q_3 - \frac 12 Q_1 +
\frac 12 (Q_8+Q_8') \label{hyper} \eeqa It is easy to check that
the $B \wedge F$ couplings for these combinations also vanish,
hence the corresponding $U(1)$'s remain in the massless spectrum.
The combination (\ref{hyper}) plays the role of hyper-charge in
the present model. The theory contains three Standard Model
families, plus one exotic chiral (but anomaly-free) set of
fields, plus multiplets with vector-like quantum numbers under
the SM group.

In this model, the quarks, leptons and Higgs fields live at
different intersections. Hence, the Yukawa couplings $Y_{ijk}$
among the Higgs fields and two fermions arise from a string world-sheet
of area $A_{ijk}$ (measured in string units) stretching between
the three intersections \cite{afiru2}, and hence $Y_{ijk} \sim
\exp(- A_{ijk} )$. Note that one family of quarks and leptons do
not have renormalizable couplings with the Higgs fields, due to
the uncancelled $Q_2$ charges, and the only chiral multiplets
which carry opposite $Q_2$ charges are charged under the weak
$SU(2)$. In addition to the above couplings, there are
generically couplings among $aa$, $ab$ and $ba$ fields, and among the
$aa$ fields. This is also the case in all the previous models of
intersecting branes in the literature). These couplings are not
exponentially suppressed. It would be interesting to examine if
these couplings may pose phenomenological challenges for these
models.

Let us emphasize that this is the first realization of a
semi-realistic spectrum in the context of fully supersymmetric
type II orientifold constructions. Clearly, there exist different
variants, obtained by changing the additional branes not directly
involved in the Standard Model  structure. The Standard Model core
structure which we have found is however relatively unique, at
least within our Ansatz for the angles on branes. It would be
interesting to explore such variants to eliminate part of the
additional vector-like matter, or the extra exotics.

\subsection{A supersymmetric GUT example}
\label{gut}

The present setup allows us to consider new possibilities in model
building. For instance, we may consider constructing a
supersymmetric grand unified model. This possibility is not
available in standard type IIB orientifolds, due to the
difficulty in getting adjoint representations to break the GUT
group to the Standard Model (see {\it e.g.}, \cite{lykken}). In
the context of intersecting branes, GUT models have been
constructed in {\it e.g.},  \cite{bgkl}, but they are
non-supersymmetric, hence suffer from a severe hierarchy problem.

In this Section we show it is extremely straightforward 
to build GUT models in the
present setup, as we illustrate with a simple four-family example.
Consider the following configuration of branes
\begin{center}
\begin{tabular}{|c|c|c|}
\hline\hline
$N_a$ & \quad  & $(n_a^1,m_a^1)\times (n_a^2,m_a^2)\times (n_a^3,m_a^3)$
\\
\hline
$10+6$ & & $(1,1)\times (1,-1) \times (1,{\widetilde {1/2}})$ \\
\hline
$16$ & & $(0,1)\times (1,0) \times (0,-{\widetilde {1}})$ \\
\hline \hline
\end{tabular}
\end{center}
which is supersymmetric for $\arctan \chi_1 - \arctan \chi_2 + \arctan
(\chi_3/2)=0$. We consider that the first set of $16$ branes is split in
two parallel stacks of $10$ and $6$. The resulting spectrum is
\beqa
& U(5)\times U(3) \times USp(16) & \nonumber \\
& 3(24+1,1,1) + 3(1,8+1,1) + 3(1,1,119+1) & \nonumber \\
& 4({\ov{10},1,1}) + (5,1,16) + 4 ({\ov 5},{\ov 3},1) + (1,3,16) +
4(1,3,1)& \eeqa where we have ignored for simplicity that one of
the $U(1)$'s is actually anomalous and massive. The model is a
four-family $SU(5)$ GUT, with additional gauge groups and matter
content. Notice that turning on suitable vev's for the adjoint
multiplets the model corresponds to splitting the $U(5)$ branes.
This provides a geometric interpretation of the GUT Higgsing to
the Standard Model group upon splitting $U(5)\to U(3)\times
U(2)\times U(1)$. Also, it provides the construction of a new
Standard Model with four quark-lepton families with correct
quantum numbers. In this framework hyper-charge is given by the
linear combination familiar in grand unification. It is important
to point out that since all the Standard Model gauge groups would
arise from branes wrapped on parallel but otherwise identical
cycles, this string construction provides a natural initial
condition for the unification of gauge couplings. Hence, such
models provide a stringy embedding of the basic philosophy in
traditional GUT.

Clearly the above model can be improved by complicating the
configuration, but we refrain from doing so. Our purpose is to
illustrate it is possible to build such GUT's with reasonable
numbers of families, and adjoint Higgs multiplets. Notice that a
generic feature of these GUT constructions is that the adjoints
are exact moduli in the model, since they are associated with the
brane motion in transverse space. This absence of Higgs
self-interactions leads, upon breaking to Standard Model gauge
group, to the matter in the adjoint representation
of the Standard Model  factors.
This is very reminiscent of what happened in heterotic string
GUT's \cite{guts}.

\section{Small instanton transitions}
\label{small}

In this Section we briefly discuss how our general class of
models with branes at angles is connected to the familiar
non-chiral $\IZ_2\times \IZ_2$ type IIB orientifold
\cite{berkooz}. In that respect, it is useful to recall the
T-duality between configurations of branes at angles and branes
carrying gauge magnetic fluxes (see {\it e.g.},  \cite{bgkl}, and
\cite{raul} for a recent discussion). In particular, a brane
wrapped at angles on cycles $(n^i,m^i)$ correspond to a brane
fully wrapped on the two-tori (in fact, multi-wrapped $\prod_i
n^i$ times), with total magnetic flux $m^i/n^i$ in the $i^{th}$
two-torus.

It follows that in the $\IZ_2\times \IZ_2$ IIA model, D6-branes
along the O6-planes are mapped to D9-, D5$_i$-branes in the
$\IZ_2\times \IZ_2$ type IIB orientifold in \cite{berkooz}. On
the other hand, our models with D6-branes at non-trivial angles
in two directions T-dualize to configurations where some
D9-branes carry fluxes in two complex directions. The
corresponding non-zero instanton numbers endow the D9-branes with
D5$_i$-brane charge (see \cite{magnetised} for discussion). Specifically,
$N$ D6-branes along {\it e.g.},
$(1,0)\times (n_2,m_2)\times (n_3,m_3)$ T-dualize to a bound
state of $Nn_2n_3$ D9-branes and $Nm_2m_3$ D5$_1$-branes.
Consequently, such models contain a smaller number of pure
D5$_i$-branes. It is clear that these models are connected to the
basic $\IZ_2\times \IZ_2$ orientifold by small instanton
transitions \cite{small}, in which some of the D5$_i$-branes are
dissolved into the D9-branes, and expand to fill the corresponding
four-torus in a uniform manner. Obviously, the intermediate steps
in the process involve non-constant self-dual field-strength gauge
configurations, which are not described by a free world-sheet CFT.
However, the final configuration, with constant flux, admits such a
description.

The fattening of small instantons can be followed using field theory
techniques, in the region of small instanton size. In fact, it is an
interesting exercise (left to the reader) to verify that in the
six-dimensional context there exist flat directions which connect the
$U(16)^2$ type IIB $\IT^4/\IZ_2$ model in \cite{bs,gp} with the models in
\cite{magnetised} \footnote{A useful hint is that e.g. in the $U(13)\times
U(4)\times U(3)$ example in \cite{magnetised}, this gauge group is
embedded in the original $U(16)^2$ as follows from the decomposition
$U(16)^2\supset U(13)\times U(4)\times U(3)^5 \supset U(13)\times
U(4)\times U(3)_D$, where $U(3)_D$ is the diagonal combination of the five
$U(3)$ factors. This breaking to the diagonal accounts for the (Landau
level) multiplicities in the final model with fluxes.}. Such flat
directions represent the effective field theory description of the small
instanton transition, analogous to the flat space discussion in
\cite{small}. Obviously, the field theory analysis is perturbative in the
vevs, and hence valid only close to the small instanton point. Hence it
cannot be used to follow the flat direction for a finite distance, namely
until the instanton has become a uniform flux.

In the picture of branes at angles, the process corresponds to recombining
D6-branes wrapping different intersecting cycles. In the intermediate
steps the recombined cycle is complicated, hence it is difficult to describe
it in detail. After flattening it out (when possible) preserving its
homology class, it corresponds to D6-branes in a cycle at non-trivial
angles. These recombinations have been considered in the
non-supersymmetric case \cite{afiru1,afiru2,ralph}, where they are
triggered by tachyons. In our present supersymmetric context, they rather
correspond to flat directions in moduli space.

Notice the striking feature that using these transitions one can
generate chiral models starting from non-chiral ones. The
situation is reminiscent of the chirality changing small
instanton transitions in \cite{ksafiuv,opp}. Consequently, one
can also use small instanton transitions to relate different
chiral models within our class, differing in their D6-brane
wrapping numbers, and with different chiral content (and/or
different gauge group). In fact, in the following we discuss a specific
Standard Model toy example where the number of families changes by
such a process.

\begin{table}[htb] \footnotesize
\renewcommand{\arraystretch}{1.25}
\begin{center}
\begin{tabular}{|c||c|c|}
Type & $N_a$ & $(n_a^1,m_a^1) \times (n_a^2,m_a^2) \times (n_a^3,m_a^3)$
\\
\hline
Spectator & 6+2 & $(1,0)\times (1,1)\times (1,-1)$ \\
          &   4 & $(1,0)\times (1,0)\times (1,0)$ \\
          &   8 & $(1,0)\times (0,1)\times (0,-1)$ \\
          &  16 & $(0,1)\times (1,0)\times (0,-1)$ \\
\hline
Before    &   4 & $(1,1)\times (1,-1)\times (1,0)$ \\
          &  12 & $(0,1)\times (0,-1)\times (1,0)$ \\
\hline
After     &   4 & $(1,2)\times (1,-2)\times (1,0)$
\end{tabular}
\end{center}
\caption{\small Wrapping numbers/fluxes for the small instanton
transition. The upper piece of the Table lists branes unaffected by the
transition. The lower pieces describe the branes existing before and after
the transition. The first set of 8 branes is split in $6+2$ to yield gauge
group $U(3)\times U(1)$.}
\label{transit}
\end{table}

Let us consider the $\IZ_2\times \IZ_2$ models, with rectangular
two-tori, described in Table~\ref{transit}. Consider the initial
model, where D6-brane wrapping numbers are given by the first six
rows in the Table. In the T-dual picture the model is understood
in terms of instanto bundles as follows. Consider the model without
magnetic fluxes in
\cite{berkooz}, with group $USp(16)^4$, the four factors arising
from D9- and D5$_i$-branes. Consider dissolving 8 D5$_1$- and 4
D5$_3$-branes (and images) as instantons within a $U(1)\times
U(1)$ sub-group of $USp(16)_9$. In the decomposition \beqa USp(16)_9 \supset
U(4)\times U(2)\times USp(4) \label{split} \eeqa
the two $U(1)$ generators
which acquire non-zero flux correspond to the $U(1)$'s within $U(4)$
and $U(2)$, respectively.
The surviving
group from the D9-branes is the commutant of the gauge
background, namely $U(4)\times U(2)\times USp(4)$. The final
spectrum can be computed directly in string theory in the flux or
angle picture, or in field theory using the index theorem (see
\cite{bachas} for a discussion). The chiral spectrum is \beqa
&"U(4)"\times U(2) \times USp(12)\times USp(4)\times USp(8) \times
USp(16) & \nonumber \\
&2\times (4,{\ov 2};1,1,1,1) + (4,1;12,1,1,1) + ({\ov 4},1;1,1,1,16) + &
\nonumber \\
& (1,{\ov 2};1,1,8,1) + (1,2;1,1,1,16) & \eeqa where the
symplectic groups arise from (a) D9-branes without flux, and (b)
D5$_i$-branes not dissolved as fat instantons. The quotation marks for
$U(4)$ denote that we are actually interested in splitting it as
$U(3)\times U(1)$ (by separating the D6-branes, or equivalently
by introducing D9-brane Wilson lines). The resulting model
corresponds to a two-family Standard-like model. One can even
obtain a sensible hyper-charge by splitting the $USp(4n)$
symplectic factors as $U(1)^n$ so as to generate additional
$U(1)$'s. In fact the linear combination \beqa Q_Y = \frac 16 Q_3
-\frac 12 Q_1 + \frac 12 (Q_{12} + Q_{4} + Q_{8}) \label{hyp1}
\eeqa (where $Q_3$, $Q_1$ arise from $"U(4)"$, and $Q_{12}$,
$Q_4$, $Q_8$ arise as diagonal combinations of the $U(1)$'s from
symplectic factors) is automatically anomaly-free and massless,
and plays the role of hyper-charge in the above model. The net
chiral content with respect to Standard Model interactions is
\beqa 2(3,2)_{1/6} + 2({\ov 3},1)_{1/3} + 2({\ov 3},1)_{-2/3} +
2(1,2)_{-1/2} + 2(1,1)_1 + 2(1,1)_0 \eeqa Namely two standard
quark-lepton generation (plus right handed neutrino).

\medskip

In the angle picture, let us consider 
the 4 D6-branes along $(1,1)\times (1,-1)\times (1,0)$ (and their
$\Omega R$ images) combine with the 12 along $(0,1)\times
(0,-1)\times (1,0)$ (and images) to give 4 D6-branes along
$(1,2)\times (1,-2)\times (1,0)$ (and images). This process is
possible since the total homology class is conserved, and is
triggered by vev's for scalars in strings stretching between the
stacks involved. The final model has D6-branes with wrapping
numbers given by the first four and the last rows in
Table~\ref{transit}.

Let us describe the process in the T-dual picture. The final model
corresponds, in terms of the underlying $USp(16)^4$, to the
following. Consider dissolving 8 D5$_1$- and 16 D5$_3$-branes
(and images) as instantons in a $U(1)\times U(1)$ sub-group
corresponding to the splitting (\ref{split}). The resulting flux
structure corresponds to the final model. Hence the transition
connecting both models amounts to dissolving 12 additional
D5$_3$-branes (and images) as instantons within the same $U(1)$
sub-group where the previous 8 were already dissolved. In
particular this shows that the unbroken D9-brane gauge group has
identical generators in both cases. However the chiral fermion
content may differ, due to the presence of additional flux
modifying the index of the Dirac operator in the internal space
(or the intersection number in the angle picture). The final
spectrum can be computed with string theory or index theory
techniques, and reads \beqa
&"U(4)"\times U(2) \times USp(4)\times USp(8) \times USp(16) &\\
&6\times (4,{\ov 2};1,1,1,1) + 2\times (4,2;1,1,1,1) + ({\ov 4},1;1,1,16)
+ & \nonumber\\
& 2\times (1,{\ov 2};1,1,8,1) + 2\times (1,2;1,1,1,16) & \eeqa
Upon splitting $U(4)\to U(3)\times U(1)$ the model corresponds to
an eight-family Standard Model. In this case hyper-charge can be
obtained by splitting the symplectic factors into $U(1)$'s and
considering the anomaly-free and massless combination \beqa Q_Y =
\frac 16 Q_3 -\frac 12 Q_1 + \frac 12 (Q_{4} + Q_{8}) \eeqa The
chiral spectrum with respect to Standard Model gauge interactions
is \beqa 8(3,2)_{1/6} + 8({\ov 3},1)_{1/3} + 8({\ov 3},1)_{-2/3}
+ 8(1,2)_{-1/2} + 8(1,1)_1 + 8(1,1)_0 \eeqa Namely, there are eight standard
quark-lepton families. Hence the process of dissolving additional
D5-branes in the gauge bundle on the D9-branes leads to phase
transitions changing the number of chiral families.

A minor difficulty in the above transition is that the hyper-charge
generator is not the same in both theories. This can be avoided
by removing the $Q_{12}$ term in (\ref{hyp1}), although this
yields to exotic hyper-charges in the initial model. Leaving these subtle
points aside, we would like to emphasize that it is
remarkable that one can describe quite explicitly these
transition in relatively realistic models.

We have succeeded in showing that small instanton transitions can mediate
changes in the number of families in a model. Moreover, the T-dual
interpretation of these processes as recombination of cycles provides a
useful tool in analyzing these transitions in the type II orientifold
setup. We believe these techniques can be useful in the study
interesting phenomena in a simple geometric setup, and hence are
complementary to other  realizations of these transitions
\cite{ksafiuv,opp}.

\section{D-term supersymmetry breaking}
\label{ssb}

As mentioned above, the general models we have considered are
supersymmetric for specific choices of the untwisted moduli $\chi_i$. In
this Section we briefly consider the main supersymmetry breaking effects
on the open string sector as one moves away from the special
supersymmetric values for $\chi_i$.

On general grounds, the complex structure moduli $\chi_i$ are expected to
couple open string modes in our D6-branes on 3-cycles (A-branes) as
Fayet-Iliopoulos (FI) terms. This is mirror to the statement that
K\"ahler moduli couple to B-branes as FI-terms, a familiar situation for
D-branes at singularities \cite{dm}. 
(Note that related techniques have been employed in the
blowing-up procedure of the type IIB orientifold singularities 
\cite{clew}).
This has appeared in a related
context in \cite{kachru}, and in situations where there are
D-branes with ($B_{NS}$ or equivalently magnetic field) fluxes in
\cite{witten}. Following the
latter, we expect the corresponding FI-terms to be proportional to the
deviation from the supersymmetric situation, in particular we expect the
terms in the effective $D=4$ action
\beqa
\sum_a \int d^4x \left( \theta^1_a + \theta^2_a + \theta^3_a \right) D_a
\label{fi}
\eeqa
where $\theta_a^i$ is understood as a function of $\chi_i$ for fixed
wrapping numbers $(n_a^i,m_a^i)$. For instance, for square tori
\beqa
\theta_a^i = \frac{m_a^i}{n_a^i} \chi_i.
\eeqa
The FI-term vanishes for the supersymmetric situation $\theta_1+\theta_2+
\theta_3=0$, hence in general it is proportional to the deviation from
this case. It is easy to see that this term reproduces the leading order
splitting between scalar and fermion masses, as one would obtain from the
string computation. Namely, in $ab$ sector, chiral fermions remain
massless at tree level,  while their scalar partners obtain a mass
proportional to
$\delta \theta= \sum_i(\theta^i_a-\theta^i_b)$.

In supersymmetric models, the familiar arguments in \cite{dsw} relate the
existence of Green-Schwarz anomaly cancellation mechanism with the
existence of FI terms, controlled by the partners of the fields mediating
the GS interactions. Their precise determination also requires knowledge
about their K\"ahler potential, which should be easily determined for the
untwisted complex structure moduli in our models. However, we skip the
derivation of the FI terms, and briefly discuss the physics arising from
Eq.(\ref{fi}).

The turning of FI terms when untwisted moduli are shifted from the
supersymmetric values actually does not automatically imply
breaking of supersymmetry. As is familiar in heterotic
constructions, some scalar fields may acquire vev's so as to make
the D-term vanish. Hence, supersymmetry would be restored in the
shifted vacuum. An important difference with respect to heterotic
models is that the FI-terms are not related to the dilaton, and
can be tuned at will by tuning the untwisted moduli
\footnote{This is analogous to the observation in \cite{iru} for
FI-terms and twisted moduli in standard type IIB orientifolds.}.
The physics behind this process is that the original D6-brane
configuration is no longer supersymmetric for the new choice of
untwisted moduli (since the angles are changed), so some
intersecting D6-branes recombine into a smooth 3-cycle which is
supersymmetric. This recombination is described by the vev
acquired by certain scalar fields at intersections.

It is an interesting question whether supersymmetry can always be
restored in this fashion. Despite the lack of a general argument, we
strongly suspect that this is the case, at least in compact models. Notice
that however non-compact models allow for supersymmetry breaking by this
FI-term mechanism (see \cite{kachru} for discussion).

In order to provide a simple illustrative example, let us
consider the string-GUT model in Section \ref{gut}, where there
is only one relevant set of angles, namely those formed by the
$U(8)$ branes with the horizontal axes, denoted by $\theta_i$
henceforth. There are two relevant kinds of scalars, those
arising at the intersections between the $U(8)$ branes and their
images, $\phi_{aa'}$, and between the former and the $USp(16)$
branes, $\phi_{ab}$. The corresponding chiral multiplets carry
opposite charges with respect to the single $U(1)$ symmetry in
the model, hence the D-term has the schematic structure \beqa D
\, = \, 2 |\phi_{aa'}|^2 - |\phi_{ab}|^2 +
(\theta_1+\theta_2+\theta_3) \eeqa Hence, for deformations such
that $\sum_i \theta_i>0$, the fields $\phi_{ab}$ become tachyonic
and acquire a vev, restoring supersymmetry. This corresponds to
recombining the $USp(16)$ and the $U(8)$ branes. For $\sum_i
\theta_i<0$, it is $\phi_{aa'}$ which acquire vev's to restore
supersymmetry, triggering the recombination of the $U(8)$ branes
and their images.

It would be interesting to explore these processes in more detail, both in
their effective field theory and in their geometric description.

\section{The Relation to Compact Singular $G_2$ Manifolds}
\label{gtwo}

In this Section we briefly outline a different (more formal)
aspect of our models\footnote{The comments in this Section lie
outside the main line in this paper. We advice readers with more
phenomenological interests to safely skip it.}. Recently there
has been a lot of interest in the study of the dynamics of
M-theory on 7-dimensional manifolds $X_7$ admitting a $G_2$
holonomy metrics \cite{gtwo,amv,aw}. The interest stems from the
fact that such compactifications lead to four-dimensional $N=1$
supersymmetric field theories, with gauge interactions determined
by the singularity structure of $X_7$. Moreover, such
constructions have provided a geometric interpretation \cite{amv}
of the duality between type IIA configurations with D6-branes on
the special lagrangian 3-cycle in the deformed conifold, and type
IIA on the resolved conifold, without D6-branes but with RR
2-form fluxes \cite{vafa}. These results have been extended in
diverse directions (see, {\it e.g.}, \cite{kot}), 
and suggest interesting connections with
gauge theory dynamics and string duality.

Topological manifolds admitting a $G_2$ metric are not easy to 
characterize, as opposed to, {\it e.g.}, spaces admitting a $SU(n)$
holonomy metric, which can be characterized by a topological condition
(the vanishing of the first Chern class). The explicit
construction of $G_2$ metrics is difficult, and has only been
achieved in a few non-compact examples constructed in
\cite{metrics} and the more recent generalizations in \cite{g2n}. (For
applications to regular configurations of M-theory with $N=1$
supersymmery, see  {\it e.g.}, \cite{g2v} and references therein.)
However, string theory duality provides a simple strategy to
obtain topological spaces which admit a $G_2$ metric, without
constructing it explicitly. Basically, any type IIA configuration
preserving $D=4$ $N=1$ supersymmetry, and including at most
D6-branes and O6-planes, will lift to an M-theory
compactification on a $G_2$ manifold (see \cite{jaume,nunez} for nice
discussions). The topological information can be used to obtain
interesting qualitative features of these theories.

In this respect, the $D=4$ $N=1$ supersymmetric type IIA orientifolds with
D6-branes and O6-planes at angles, studied in \cite{blumen4d,bonn1} and in
this paper, correspond to M-theory compactifications on $G_2$ manifolds.
In configurations where the RR 7-form charges are locally cancelled
(namely, 2 D6-branes and 2 images on top of each O6-plane in the
configuration), the M-theory lift
is remarkably simple. The M-theory circle is constant over the base space
$B_6$, leading to a total variety $(B_6\times S^1)/\IZ_2$, where the
$\IZ_2$ flips the coordinate parametrizing the M-theory circle, and acts
on $B_6$ as an antiholomorphic involution (hence changing the holomorphic
3-form to its conjugate). This is analogous to the discussion in
\cite{kmg}.

Unfortunately, such models lead generically to non-chiral spectra, in the
sense that even though one obtains chiral multiplets, they arise in real
representations of the gauge group. On the other hand,
configurations with D6-branes away or
not fully aligned with the O6-planes would led to more
involved M-theory lifts. However, they are of great interest since, as
shown in our general constructions, they lead to chiral gauge theories.

Certainly, there exist simpler lines of attack to obtain chirality out of
$G_2$ singular spaces. In particular, the simplest type IIA supersymmetric
configuration leading to chiral fermions is simply two intersecting
D6-branes in flat space, related by an $SU(3)$ rotation. Its M-theory lift
would correspond to a rigid 7-dimensional $G_2$ singularity, and is the
basic building block for engineering chiral theories using $G_2$
geometries; the chiral multiplet arises from M2-brane wrapping a collapsed
two-cycle. In this sense, our configurations are more
complicated, and what they provide is a consistent embedding of this
building block singularity in a compact setup.

We expect that the generic class of models described here may exhibit
some interesting phenomena in this context, in particular the existence
of non-perturbative equivalences among seemingly different models, which
nonetheless share the same M-theory lift, in analogy with \cite{kmg}.
On the other hand the type IIA transitions in which intersecting D6-branes
recombine (the T-dual of the small instanton transitions) would have
interesting M-theory descriptions, in which the topology of the $G_2$
space changes. It would be interesting to explore possible connections of
such process with \cite{amv,pioline}. We hope that our explicit
constructions may provide a useful laboratory to probe new ideas in this
exciting development.

\section{Conclusions}
\label{conc}

In this paper, we constructed four-dimensional $N=1$ supersymmetric type
II orientifolds with branes at angles. We provided the first D-brane
construction of a three-family $N=1$ supersymmetric vacuum solution with
the Standard Model gauge  group $SU(3)_C \times SU(2)_L \times U(1)_Y$ as part
of the gauge structure. We have also illustrated other possibilities for
model building by constructing a supersymmetric GUT model with (four) chiral
families and adjoint Higgs multiplets, the first example of its kind in
the orientifold setup. Although we have not discussed here,
it is quite straightforward to construct other
extensions of the Standard Model, such as the left-right symmetric models
({\it e.g.}, the Pati-Salam type).

Even though our models are explicit string realizations of the
brane-world scenario, they generically require a high string
scale. In models with intersecting D6-branes, the experimental
bounds on masses of Kaluza-Klein replicas of Standard Model gauge
bosons imply that the internal dimension cannot be large (since
there is no dimension transverse to {\em all} Standard Model
branes). Hence, a large Planck mass can be generated only from a
large string scale, and not from a large volume. Specifically,
one obtains \beqa g_{YM}^2 M_P^{(4d)} = M_s \frac{\sqrt{V_6}}{V_3}
\eeqa where $V_3$ is the volume of the cycle wrapped by the
corresponding brane, and $V_6$ is the total internal volume.
Moreover, large anisotropies in the internal space would
generically reflect in  different gauge couplings for different
gauge group factors. For nearly isotropic configurations, the
string scale is of the order of the Planck scale. There is
however more freedom than in the traditional heterotic approach,
and it could be used to lower the string scale to, {\it e.g.},
$10^{16}$ GeV, a certainly desirable choice for GUT models.

Another interesting feature of this class of models with branes
at angles is the structure of the Yukawa couplings. Since the
quarks, leptons and Higgs fields are located at different
intersections of the branes, the Yukawa couplings $Y_{ijk}$ are
generically exponentially suppressed by the area $A_{ijk}$ of the
string world-sheet stretching between the locations of these
fields (measured in string units) \cite{afiru2}, {\it i.e.},
\begin{equation}
Y_{ijk} \sim \exp(-A_{ijk})
\end{equation}
These exponential factors may provide an interesting geometrical
explanation for the observed fermion masses. In order for the
Yukawa couplings not to be negligibly small, the area of the
string world-sheet (which is typically the compactification scale)
cannot be much larger than the string length, so internal
dimensions should be of the order of the string scale. Note that
in addition to the above Yukawa couplings, there are generically
couplings among $aa$, $ab$ and $ba$ fields, and among $aa$
fields. These couplings are not exponentially suppressed.
Therefore, in studying the resulting phenomenology, one should
examine whether they may pose phenomenologically challenges to
this general class of models involving branes at angles.
Hopefully, they might be useful, {\it e.g.}, to get rid of the
unwanted non-chiral matter in the $aa$ sector, even though we
have no concrete proposal in this respect.
We also note in passing that unlike the recent models in \cite{bklo},
our models have absolute proton stability
due to symmetries (as discussed in \cite{afiru2}).

The basic model building rules that we have constructed, allow
for the exploration of a potentially large class of supersymmetric
models. In particular, to simplify the conditions from
supersymmetry, we have mainly restricted our search to D6-branes
at angles of the form $(\theta_1,\theta_2,0)$,
$(\theta_1,0,\theta_3)$ or $(0,\theta_2,\theta_3)$. It is quite
remarkable that within this restricted class of models, there
exist three-family Standard-like models. The three-family model
is however not fully realistic, as it contains extra vector-like
multiplets as well as exotic chiral matter. Nevertheless, it is
possible that variants of this (or other) model(s) may eliminate
these additional states, and lead to solutions closer to the
Standard Model (in particular, it would be interesting to
reproduce the very economical spectrum structure in \cite{imr}).
Clearly, a detailed search of realistic models deserves further
investigation.

In this regard, it is interesting to note that supersymmetric
D6-branes with three non-trivial angles, say $(n_i,m_i)=(1,1)
\times (1,1) \times (1,-1)$, contribute to some tadpole
conditions (but not all!) with the same sign as that of the
O6-planes. This implies that even though the configuration is
supersymmetric, the branes can carry negative RR charges under
some RR forms in the dimensional reduction on the internal space.
This allows for more flexibility in satisfying the tadpole
conditions, and may give more room for embedding of the Standard
Model.

The general class of models with branes at angles are connected to familiar
orientifolds by (the T-dual of) small instanton transitions. For instance,
the three-family model we presented is connected by such a transition to
the non-chiral $\IZ_2 \times \IZ_2$ orientifold in \cite{berkooz}. In the
picture of D6-branes at angles, the transition amounts to a recombination
of the intersecting cycles, resulting in D6-branes not fully
aligned with the O6-planes, and leading to chiral matter. The recombination
description provides a simple setup to analyze these transitions, and
their field theory interpretation in detail.

We have also discussed the main physical effects when the
untwisted moduli are chosen away from the supersymmetric point.
The model develops FI terms proportional to the deviation from
the supersymmetric situation. The corresponding non-zero D-terms
in general force some scalars to acquire vev's and restore
supersymmetry at the restabilized vacuum. This is the field
theory counterpart of a process in which the intersecting
D6-branes, non-supersymmetric in the new situation, recombine and
wrap a new 3-cycle which is supersymmetric in the new, deformed,
complex structure.

As this general class of supersymmetric orientifold models involve only
D6-branes and O6-planes, their M-theory lift  correspond to
compactifications on purely geometrical background, in fact, a compact,
singular
7-manifold with $G_2$ holonomy. Given the recent interest in M-theory
compactifications on such spaces, we expect our general class of
orientifold models may lead to new insights into the construction of spaces
with special holonomy 
leading to four-dimensional gauge theories
with chiral fermions, and into new physical phenomena in such
compactifications.

There are many promising avenues to explore in supersymmetric
orientifolds with D6-branes at angles, both from the phenomenological and
the theoretical viewpoints. We hope our results here have provided the
first steps in some of these directions.

\acknowledgments

We thank Gerardo Aldazabal, Savas Dimopoulos, Jens Erler,
Gary Gibbons, Jaume Gomis, Luis Ib\'a\~nez, Paul
Langacker, Hong L\"u, Chris Pope, Raul Rabad\'an and Edward Witten 
for useful discussions. G.S. thanks
Ralph Blumenhagen and Boris K\"ors for email correspondence, and
Matt Strassler for sharing his insights on generation-changing
transitions. We would like to thank the Theory Division at CERN
(M.C. and G.S.), CAMTP, University of Maribor, Slovenia (M.C.)
Amsterdam workshop on String theory (M.C.) and Benasque workshop (M.C. and
A.M.U.) for hospitality during
the course of the work. A.M.U. thanks M.~Gonz\'alez for kind
encouragement and support. This work was supported in part by
U.S.\ Department of Energy Grant No.~DOE-EY-76-02-3071 (M.C.), in
part by the Class of 1965 Endowed Term Chair (M.C.), UPenn SAS
Dean's funds (G.S.) and the NATO Linkage grant 97061 (M.C.).

\clearpage

\appendix

\section{R-R tadpole and Supersymmetry conditions}\label{tadpoles}

In this Appendix we  derive the  tadpole cancellation conditions for the
$\IZ_2\times\IZ_2$ orientifold model of the type discussed in the main
text. The results can be generalized in a rather straightforward way to
other types of $\IZ_N$ orientifold models.

\begin{figure}
\begin{center}
\centering
\epsfysize=3.5cm
\leavevmode
\epsfbox{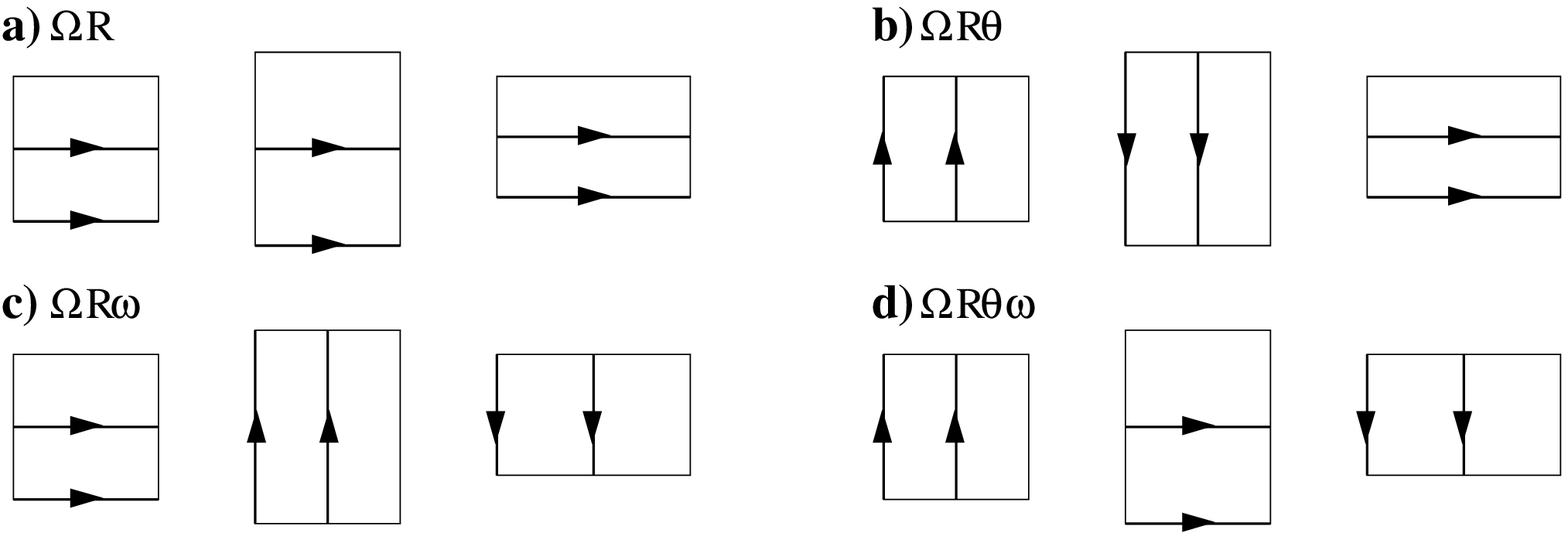}
\end{center}
\caption[]{\small O6-planes in the orientifold of $\IT^6/(\IZ_2\times
\IZ_2)$.}
\label{orient}
\end{figure}

As obtained in \cite{blumen6d,blumen4d,bonn1}, $\Omega R$ orientifolds of
type IIA toroidal orbifolds
do not contain twisted crosscap tadpoles. Correspondingly, D-branes
wrapped on factorized three-cycles on the six-torus (i.e. wrapped on
one-cycles in each complex plane) do not generate twisted disk tadpoles.
Therefore, only cancellation of untwisted RR tadpoles should be imposed,
namely, the consistency conditions are the same as for a set of
O6-planes and D6-branes (and their images) in the six-torus. This is
simply Gauss law, the cancellation of the total charge under the RR
7-form. Such charges are proportional to the homology class of the
wrapped three-cycles. Let us consider the case of rectangular two-tori,
and denote $[a_i]$, $[b_i]$ the $(1,0)$ and $(0,1)$ homology one-cycles in
the $i^{th}$ two-torus. The O6-planes fixed under the orientifold actions
$\Omega R$, $\Omega R\theta$, $\Omega R \omega$, $\Omega R\theta\omega$,
shown in figure~\ref{orient}, carry an overall charge proportional to
\beqa
-4\times 8 \times (\; [a_1] \times [a_2] \times [a_3]
- [b_1] \times [b_2] \times [a_3] - [a_1] \times [b_2] \times [b_3] -
[b_1] \times [a_2] \times [b_3] \; )
\label{chor}
\eeqa
where the $-4$ is the charge of a single O6-plane (in D6-brane charge
units, and as counted in the covering space), and 8 is the number of
O6-planes of each kind.

We would like to cancel this charge by introducing sets of $N_a$ D6-branes
wrapped on the three-cycle defined by wrapping numbers $(n_a^i,m_a^i)$,
and their orientifold images, with wrapping numbers $(n_a^i,-m_a^i)$. The
total D6-brane charge is
\beqa
\sum_a N_a \prod_{i=1}^3 (n_a^i [a_i] + m_a^i [b_i])
+\sum_a N_a \prod_{i=1}^3 (n_a^i [a_i] - m_a^i [b_i])
\label{chdb}
\eeqa
The requirement that charges (\ref{chor}), (\ref{chdb}) add up to zero
yields the RR tadpole constraints
\beqa
\sum_a N_a n_a^1 n_a^2 n_a^3 - 16 & = & 0 \nonumber \\
\sum_a N_a n_a^1 m_a^2 m_a^3 + 16 & = & 0 \nonumber \\
\sum_a N_a m_a^1 n_a^2 m_a^3 + 16 & = & 0 \nonumber \\
\sum_a N_a m_a^1 m_a^2 n_a^3 + 16 & = & 0
\label{tadpole1}
\eeqa

\medskip

The procedure carries over in the same spirit for the case with some
tilted tori. For instance, consider only the third torus is tilted,
\begin{figure}
\begin{center}
\centering
\epsfysize=3.5cm
\leavevmode
\epsfbox{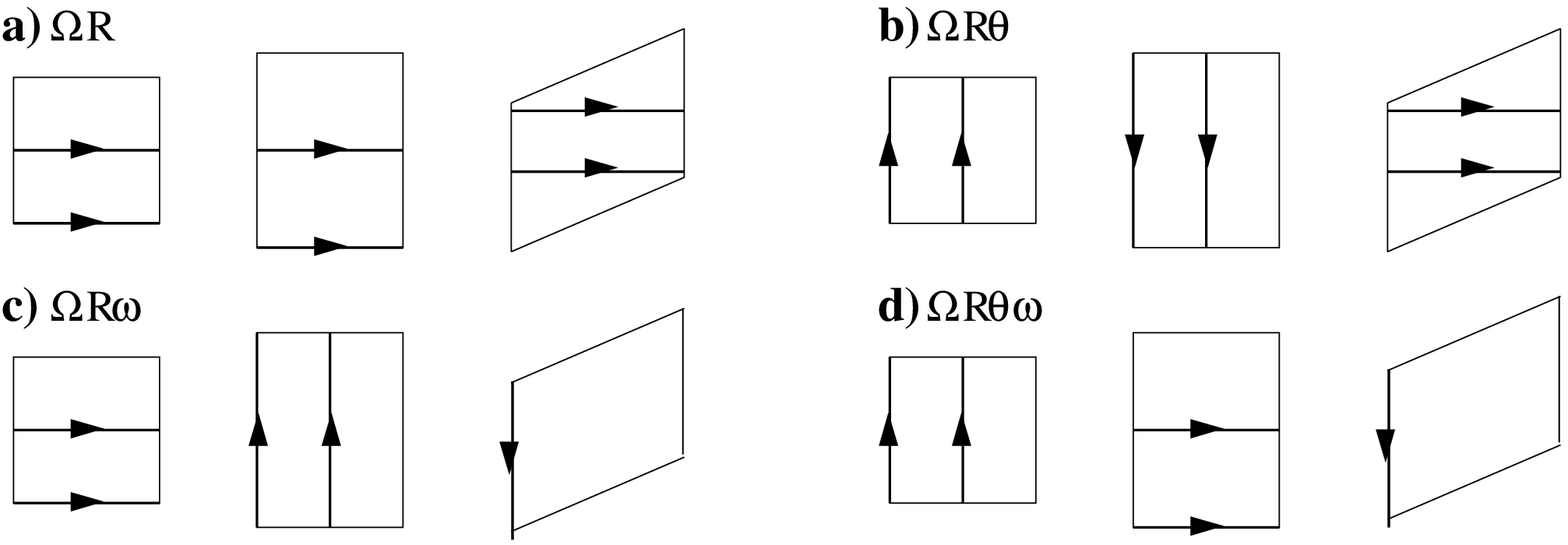}
\end{center}
\caption[]{\small O6-planes in the orientifold of $\IT^6/(\IZ_2\times
\IZ_2)$ where the third two-tori is tilted.}
\label{tiltorient}
\end{figure}
the total O6-plane RR charge is
\beqa
-4 \times 4 \times \{ [a_1]\times [a_2]\times (2[a_3]-[b_3]) &-& [b_1]\times
[b_2]\times (2[a_3]-[b_3]) \nonumber \\
&-& [a_1]\times [b_2]\times[b_3] - [b_a]\times
[a_2]\times [b_3] \}
\eeqa
where we have taken into account that due to the tilt there are only four
O6-planes of each kind, see figure \ref{tiltorient}. 
The charge of sets of $N_a$ branes with wrapping numbers $(n_a^i,m_a^i)$,
and their images, with wrapping numbers $(n_a^i,-m_a^i-n_a^i)$ are
\beqa
\sum_a N_a \prod_{i=1}^3 (n_a^i [a_i] + m_a^i [b_i])
+\sum_a N_a \prod_{i=1}^3 (n_a^i [a_i] - (m_a^i+n_a^i) [b_i])
\eeqa
Cancellation of RR charges leads to
\beqa
\sum_a N_a n_a^1 n_a^2 n_a^3 - 16 & = & 0 \nonumber \\
\sum_a N_a n_a^1 m_a^2 {\tilde m}_a^3 + 8 & = & 0 \nonumber \\
\sum_a N_a m_a^1 n_a^2 {\tilde m}_a^3 + 8 & = & 0 \nonumber \\
\sum_a N_a m_a^1 m_a^2 n_a^3 + 16 & = & 0
\label{tadpole2}
\eeqa
where we have defined ${\tilde m}_a^i=m_a^i+\frac 12 n_a^i$.
The general rule is to modify tadpole conditions involving the $m_a^i$ if
the $i^{th}$ two-torus is tilted. The modification is simply to replace
$m_a^i\to {\tilde m}_a^i$, and cut by half the corresponding crosscap
contribution.

\section{Cancellation of mixed gravitational anomalies}\label{gravanomalies}

We first consider the case of $\Omega R$ orientifolds of type IIA on
$\IT^6$ \cite{bgkl}, and turn to the case with additional $\IZ_2\times \IZ_2$
orbifold projections towards the end.

Consider the $\Omega R$ orientifold of type IIA on $\IT^6$. For simplicity
we center on a six-torus factorizable into three rectangular two-tori.
Other cases are worked out analogously.
In order to cancel the tadpoles, we introduce sets of $N_a$ D6-branes
wrapped on 3-cycles in the homology class $[\Pi_a]$, defined by the
wrapping numbers $(n_a^i,m_a^i)$. We also introduce their $\Omega R$
images, namely $N_a$ branes on cycles $[\Pi_{a'}]$, defined by wrapping
numbers $(n_a^i,-m_a^i)$. For convenience, we define $[\Pi_{O6}]$ the
homology class associated to the 3-cycles with wrapping numbers $(1,0)$
along each two-torus.

The RR tadpole cancellation conditions amount to cancellation of RR charge
in homology, namely
\beqa
\sum_a \,N_a\, [\Pi_a]\, +\, \sum_{a'}\, N_a\, [\Pi_{a'}]\, -\, 32\,
[\Pi_{O6}]\, =\, 0
\eeqa
The spectrum of chiral fermions is obtained from \cite{bgkl}, and for
branes on generic cycles reads
\beq
\begin{array}{ccc}
{\rm {\bf Multiplicity}} & \quad\quad &   {\rm {\bf Representation}} \\
{ [} \Pi_a ] \cdot [ \Pi_b ] & & (\fund_a,\antifund_b) \\
{ [} \Pi_a ] \cdot [ \Pi_{b'} ] & & (\fund_a,\fund_b) \\
\frac 12 [ \Pi_a ]\cdot [ \Pi_{a'} ] +4\, [ \Pi_a ]\cdot [ \Pi_{O6} ] & &
\Yasymm_a \\
\frac 12 [ \Pi_a ]\cdot [ \Pi_{a'} ] -4\, [ \Pi_a ]\cdot [ \Pi_{O6} ] & &
\Ysymm_a
\end{array}
\eeq where representations are with respect to the gauge group
$\prod_a U(N_a)$. It is a straightforward computation to check
that the cubic non-Abelian anomaly vanishes, upon use of the RR
tadpole condition.

As discussed in \cite{afiru1} in the absence of the orientifold 
projections,
the mixed $U(1)_a$-$SU(N_b)^2$ anomaly does not vanish in general, rather
it is proportional to
\beqa
A_{ab}\, =\, \frac 12\, N_a\, [\Pi_a]\cdot (\, [\Pi_b]+ [\Pi_{b'}]\,)
\eeqa
In the toroidal models in \cite{afiru1} the mixed $U(1)_a$-gravitational
anomalies vanishes automatically. This is not true in general for $\Omega
R$ orientifolded models, where the anomaly is proportional to
\beqa
A_a^{grav}\, =\, 24\, N_a\, [\Pi_a]\cdot [\Pi_{O6}]
\eeqa
Hence, for any model with branes intersecting with the O6-plane the
anomaly does not vanish, which is the case for the examples in \cite{bgkl}
(notice that in the orientifold models in \cite{imr}
gravitational anomalies vanish due to the specific choice of D6-branes,
which never intersect the O6-planes).

These anomalies are canceled by a Green-Schwarz mechanism
mediated by untwisted RR fields, as discussed in \cite{afiru1}
for mixed non-Abelian anomalies. Our treatment of the
gravitational anomalies below is novel.

Expanding the Chern-Simons couplings for the D6-branes (and the
images) and O6-planes (see {\it e.g.},  \cite{ssb}) \beqa
\int_{D6}\, {\cal C}\, e^{F}\, {\sqrt{{\hat A}(R)}} \quad \quad;
\quad\quad \int_{O6}\, {\cal C}\, {\sqrt{{\hat L}(R)}} \eeqa we
obtain the following relevant interactions \beqa & \frac 12
\int_{D6_a} C_3\wedge \tr (F_a \wedge F_a) \quad & ;  \quad
\int_{D6_a} C_5 \wedge \tr F_a \nonumber \\
& - \int_{D6_a} C_3\wedge \tr (R \wedge R) \quad & ; \quad
(-4)\times \frac 12 \int_{O6} C_3\wedge \tr (R \wedge R) \quad
\label{couplsix}
\eeqa
Operating as in \cite{afiru1}, we introduce two dual basis of homology
3-cycles, $\{ [\Sigma_i] \}$, $\{[\Lambda_i] \}$, satisfying $[\Lambda_i]
\cdot [\Sigma_j] = \delta_{ij}$, and introduce the expansions
\beqa
& [ \Pi_a ] = \sum_{i} r_{ai} [ \Sigma_i ] \quad & ; \quad
[ \Pi_a ] = \sum_{i} p_{ai} [ \Lambda_i ] \nonumber \\
& [\Pi_{O6}] = \sum_i r_i [\Sigma_i] \quad & ; \quad
[\Pi_{O6} ] = \sum_{i} p_{i} [ \Lambda_i ]
\eeqa
We define the untwisted RR fields $\Phi_i = \int_{[\Lambda_i]} C_3 \quad
; \quad B_2^i = \int_{[\Sigma_i]} C_5$, Hodge duals in four dimensions.
The four-dimensional couplings read
\beqa
& \frac 12 \sum_i p_{ai} \int_{M_4} \Phi_i\, \tr (F_a \wedge F_a)\quad &
; \quad N_a \sum_{i} r_{ai} \int_{M_4} B_2^i \wedge \tr F_a \nonumber \\
& - N_a \sum_i p_{ai} \int_{M_4} \Phi_i\, \tr (R \wedge R)\quad &
; \quad \frac 12 \times (-32) \sum_{i} p_{i} \int_{M_4} \Phi_i\,
\tr (R \wedge R) \eeqa These couplings can be combined in GS
diagrams where $U(1)_a$ couples to the $i^{th}$ untwisted field,
which then couples to either two non-Abelian gauge bosons or two
gravitons, and hence may cancel both kinds of mixed anomalies.
The coefficients of these amplitudes, taking into account the
coupling from $\Omega R$ image branes is, for the mixed
non-Abelian anomaly \beqa \frac 12 \sum_i N_a  (\, r_{ai}p_{bi} +
r_{ai}p_{b'\,i}- r_{a'\,i}p_{bi}-r_{a'\,i}p_{b'\,i}\, )= N_a
[\Pi_a]\cdot [\Pi_b] + N_a  [\Pi_a]\cdot [\Pi_{b'}] \eeqa For the
mixed gravitational anomaly we have \beqa -\sum_i N_a N_b (\,
r_{ai}p_{bi} +
r_{ai}p_{b'\,i}-r_{a'\,i}p_{bi}-r_{a'\,i}p_{b'\,i}\, )
+(-32)/2 \times \sum_i N_a (r_{ai} p_i - r_{a'i} p_i  \nonumber \\
=N_a [\Pi_a]\cdot (-2\sum_b N_b [\Pi_b] - 2\sum_b' N_b [\Pi_{b'}] +
(-32) [\Pi_{O6}] ) = -2 \times 48\times N_a [\Pi_a]\cdot [\Pi_{O6}]
\eeqa
where we have used the tadpole cancellation conditions. The final
expressions, modulo numerical factors not computed carefully, have
precisely the form required to cancel the residual anomaly.

\medskip

The cancellation of mixed non-Abelian and mixed gravitational
anomalies in the $\IZ_2\times \IZ_2$ orbifold model works
analogously. In fact, considering the chiral spectrum given in
Table \ref{matter}, one can reproduce step by step the above
computation and reach analogous results. Namely the anomalies are
canceled by a Green-Schwarz mechanism mediated by untwisted RR
fields, whose couplings to gauge bosons and gravitons follow from
the Chern-Simons interactions for D6-branes and O6-planes. It is
also clear that an analogous mechanism will be at work in models
with other orbifold groups.

\end{document}